\documentclass[aps,preprint,superscriptaddress]{revtex4-1}

\usepackage{algorithm, algorithmic}
\usepackage{amsmath,amsthm,amsfonts}
\usepackage{color}

\usepackage{cases}

\usepackage{graphicx}
\usepackage{epstopdf}
\usepackage{rotating}

\usepackage{multirow}

\usepackage{subfigure}

\begin{document}
\title[A convection-diffusion model for gang territoriality]{A Convection-Diffusion Model for Gang Territoriality}

\author{Abdulaziz Alsenafi}
\email[]{Alsenafi@sci.kuniv.edu.kw}
\affiliation{Kuwait University, Faculty of Science, Department of Mathematics, P.O. Box 5969, Safat -13060, Kuwait}
\author{Alethea B. T. Barbaro}
\email[]{abb71@case.edu (corresponding author)}
\affiliation{Department of Mathematics, Case Western Reserve University, 10900 Euclid Ave, Yost Hall, Cleveland, OH 44106, USA}

\date{\today}

\begin{abstract}
We present an agent-based model to simulate gang territorial development motivated by graffiti marking on a two-dimensional discrete lattice. For simplicity, we assume that there are two rival gangs present, and they compete for territory. In this model, agents represent gang members and move according to a biased random walk, adding graffiti with some probability as they move and preferentially avoiding the other gang's graffiti. All agent interactions are indirect, with the interactions occurring through the graffiti field. We show numerically that as parameters vary, a phase transition occurs between a well-mixed state and a well-segregated state. The numerical results show that system mass, decay rate and graffiti rate influence the critical parameter. From the discrete model, we derive a continuum system of convection-diffusion equations for territorial development. Using the continuum equations, we perform a linear stability analysis to determine the stability of the equilibrium solutions and we find that we can determine the precise location of the phase transition in parameter space as a function of the system mass and the graffiti creation and decay rates.

\noindent{\bf Keywords:} agent-based model, phase transition, segregation model, crime modeling
\end{abstract}

\pacs{dsaf}
\keywords{agent-based model, phase transition, segregation model, crime modeling}

\maketitle

%% ----------------------------------------------------------------

\section{\label{S:introduction}Introduction} 
Many of the world's countries face violence related to gangs. In the United States of America alone, it is estimated that there are 1.4 million gang members, and gangs are responsible for 48\% of violence in most jurisdictions and up to 90\% in some jurisdictions \cite{fbi2011}. Gangs identify themselves by distinctive graffiti, clothing and handshakes \cite{brown1978graffiti,D2008}. In many regions such as Los Angeles, California and Phoenix, Arizona, it has been found that gangs claim territory through graffiti markings \cite{AW1997,ley1974urban}.  Because of the widespread nature and societal impact, it is an important sociological question to understand how gangs form and operate. 

Recently, the physics and mathematics community has taken an interest in crime modeling. Much of this research was influenced by Schelling's seminal work on segregation dynamics \cite{schelling1969models, schelling1971dynamic, schelling2006micromotives}; there is widespread interest in modeling social segregation, especially since many of these models can be viewed from a physical or mathematical perspective \cite{stauffer2007ising,zhang2011tipping,BCDM2013,rodriguez2016exploring}. Game theory has been widely used to study population-level effects of criminality \cite{szolnoki2009topology,short2010cooperation,szolnoki2013correlation}, sometimes including a spatio-temporal or agent-based framework into the game \cite{perc2013understanding,bruni2013if}. Predator-prey-type dynamical systems have also been used in these population-level models \cite{nuno2008triangle}.  

Other methodologies in crime modeling connect the models to data, such as clustering methods to deduce community affiliations among gang members in Los Angeles \cite{van2013community}, self-exciting point processes to study the temporal patterns of residential burglary in Los Angeles \cite{mohler2011self}, scaling laws for homicide in Brazilian cities \cite{alves2013scaling}, and an epidemiological model for the 2005 rioting in France  \cite{bonnasse2018epidemiological}.  Several agent-based models for burglary and gang dynamics are discussed in further detail below \cite{SDPTBBC2008,JBC2010,SBB2010,jones2010statistical,SBBTV2012}. In fact, crime modeling has been so effective that crime modeling has intersected with a new area, predictive policing, and much of the literature is now intertwined.  Many of these models have proven effective at predicting violence and at geographically profiling offenders \cite{mohler2012geographic,mohler2014marked,mohler2015randomized} and there are many applications of crime modeling to assisting with police districting  \cite{camacho2015decision, camacho2015multi, liberatore2016comparison}. For a more thorough overview of the literature, the interested reader is directed to \cite{d2015statistical} and \cite{castellano2009statistical}.

In a seminal paper on crime modeling \cite{SDPTBBC2008}, Short \emph{et al.} created a lattice model for burglary based on the assumption that burglars often return to homes that have previously been successfully burglarized or ones that are close to it, a well-known phenomenon often referred as the `broken windows' effect \cite{wilson1982broken}. They derived a continuum system from the discrete model, consisting of two coupled reaction-diffusion equations that describe the spatio-temporal evolution of agents and attractiveness densities, and identified under which conditions crime hot spots occur. Several agent-based models have been proposed to analyze these crime hot spots \cite{JBC2010, SBB2010}. Related models have been developed for policing such hotspots \cite{jones2010statistical}, and for the mathematical analysis of the continuum model \cite{rodriguez2010local}.

Our work considers the formation of gang territories due to graffiti markings, hearkening back to the mathematical ecology literature, where researchers first discovered the role of scent marking in territorial development for wolf packs \cite{mech1991way}.  Researchers then created models for the territorial behavior of coyotes and wolves using scent marking \cite{LWM1997,  WLM1996}. Many of these models include outside information such as a home den or information about the terrain \cite{MLC1999, MLC2006}. In some of these models, the authors give both discrete and continuum versions of their model \cite{BLP2002}.  While similar in spirit to the model developed and analyzed here, the main focus of these papers is on simulating the real-world territorial dynamics of the coyotes and wolves.  Therefore, the assumptions made in these models and the continuum limits are significantly different from those presented here.

Researchers in crime modeling have considered graffiti markings instead of scent marking, and have applied these same ideas to gang dynamics. In \cite{SBBTV2012}, Smith \emph{et al.} develop a model to describe the equilibrium densities of gangs and their graffiti in the policing division of Hollenbeck by combining the wolf and coyote models \cite{LWM1997,  WLM1996, MLC1999, MLC2006} with the biased L\'{e}vy walk with networks model by Hegemann \emph{et al.} \cite{hegemann2011geographical}. Taking a different approach, in \cite{BCDM2013}, Barbaro \emph{et al.} use a two-dimensional spin model to examine the problem of the development and formation of gang territory based on graffiti, proving that the system undergoes a phase transition.  Like \cite{BCDM2013}, our model is premised on the fact that gangs avoid rival gangs' graffiti and put down graffiti of their own as they move.  However, unlike any previous agent-based models known to the authors, in this paper we consider the temporal evolution of both agent and graffiti densities where agent dynamics are designed to follow observed behavior of gang members, allowing the results to be much more understandable and applicable in a criminological setting.

Our paper is organized as follows: in Sec.~\ref{S:discrete}, we present the agent-based model that provides a basis for the rest of the paper.  In Sec. \ref{S:phases}, we examine the different phases that this discrete model exhibits and offer an order parameter to aid in the analysis of the phase transition.  In Sec.~\ref{S:simulations}, we present simulations of the discrete model and numerically explore the phase transition.  In Sec. \ref{S:derivation}, we formally derive a set of four continuum equations from the discrete model.  In Sec. \ref{S:analyzing}, we find a steady-state for the continuum model and perform a stability analysis to find where the well-mixed solution loses stability; we then compare this with the numerically computed critical parameter values from Sec. \ref{S:simulations}.  In Sec. \ref{S:conclusion}, we conclude with a discussion of the main results of this work.

%%------------------------------------------------------
\section{\label{S:TandC}Theory and Calculations}
\subsection{\label{S:discrete}Discrete Model}

We begin with a two-dimensional $L \times L$ square lattice denoted by $S$. We assume that there are two gangs, red and blue, denoted by $A$ and $B$; we further assume that the number of agents belonging to the red and blue gangs are equal and are denoted by $N_A$ and $N_B$, respectively. The total number of agents in the system henceforth will be denoted by $N$, so that $N = N_A + N_B$. Initially, the agent locations are randomly distributed on $S$  using the multivariate uniform distribution.

In this model, multiple agents of any color can occupy the same location. We denote the number of red and blue agents at a site $(x,y)$ at time $t$ by $n_A(x,y,t)$ and $n_B(x,y,t)$, respectively. Agents add graffiti markings of their own color with some probability, and they move to avoid graffiti of the opposing color.  The amount of red graffiti is denoted $g_A(x,y,t)$ and the blue graffiti by $g_B(x,y,t)$. Initially, we assume that the lattice is devoid of all graffiti. We assume periodic boundary conditions throughout. The precise description of the model is given below.

Gang members prefer to be in territory occupied by their own gang, and avoid rivals' territories except under exceptional circumstances \cite{ley1974urban}.  Therefore, in our model, agents preferentially avoid the opposing color graffiti, performing a biased random walk once there is graffiti on the lattice. We assume that all agents move at each time step to one of their four neighboring lattice sites, so that an agent at site $(x,y)$ must move to one of the four sites $\{(x+l,y),(x-l,y),(x,y+l),(x,y-l)\}$, where $l$ is the lattice spacing. We assign the probability of a red agent to move from site $s_1=(x_1,y_1) \in S$ to a neighboring site $s_2=(x_2, y_2) \in S$ to be
\begin{align}
M_A(x_1 \rightarrow x_2, y_1 \rightarrow y_2, t)  :&= \frac{e^{-\frac{\beta}{l^2} g_B(x_2, y_2,t)}}{\sum \limits_{(\tilde x, \tilde y) \sim(x_1,y_1)}e^{-\frac{\beta}{l^2} g_B(\tilde x, \tilde y, t)}}, \label{D1:probability_agent_moves_A}
\end{align}
\noindent where $\beta$ is parameter that controls the strength of avoidance of blue graffiti and $(\tilde x, \tilde y) \sim (x_1,y_1)$ denotes the four neighbors of site $(x_1,y_1)$; $M_B$ is defined similarly, with $g_B$ changed for $g_A$.  Considering the amount of graffiti belonging to the red and blue gangs as densities $\xi_A = \frac{g_A}{l^2}$ and $\xi_B = \frac{g_B}{l^2}$ allows us to reformulate \eqref{D1:probability_agent_moves_A} as follows:
\begin{align}
M_A(x_1 \rightarrow x_2, y_1 \rightarrow y_2, t) &= \frac{e^{-\beta \xi_B(x_2, y_2,t)}}{\sum \limits_{(\tilde x, \tilde y) \sim(x_1,y_1)}e^{-\beta \xi_B(\tilde x, \tilde y, t)}}. \label{D:probability_agent_moves_A}
\end{align}
Since our model assumes that all agents must move at every time step, it follows trivially that $$\sum \limits_{(\tilde x, \tilde y)\sim (x,y)}   M_A(x \rightarrow \tilde x, y \rightarrow \tilde y, t)  = 1.$$

After the agents have moved,  the expected number of agents at site $(x, y) \in S$ is
{\footnotesize
\begin{equation*}
\begin{split}
n_A(x,y,t + \delta t) = &n_A(x,y,t) + \sum_{ (\tilde x,\tilde y) \sim ( x, y)}  n_A(\tilde x,\tilde y, t) M_A( \tilde x \rightarrow x,  \tilde y \rightarrow y, t)  \\
&-  n_A(x, y, t)\sum_{ (\tilde x,\tilde y) \sim (x,y)}  M_A(x  \rightarrow  \tilde x, y  \rightarrow  \tilde y, t),
\end{split}\label{E:number_of_agents_equation}
\end{equation*}
}
\noindent where the first sum represents the agents arriving at site $(x,y)$ and the second sum represents the agents leaving site $(x,y)$.  Similarly to the graffiti, we convert the number of red agents $n_A$
into a density $\rho_A = n_A/l^2$, which brings us to our discrete update rule for the agent density:
{\footnotesize
\begin{align}
\rho_A(x,y,t + \delta t) = &\rho_A(x,y,t) + \sum_{ (\tilde x,\tilde y) \sim ( x, y)}  \rho_A(\tilde x,\tilde y, t) M_A( \tilde x \rightarrow x,  \tilde y \rightarrow y, t) \notag \\
&-  \rho_A(x, y, t)\sum_{ (\tilde x,\tilde y) \sim (x,y)}  M_A(x  \rightarrow  \tilde x, y  \rightarrow  \tilde y, t).  \label{D:discrete_agents_A}
\end{align}
}
The update rule for the density of the blue agents is defined analogously.

As the agents move, they add graffiti to the lattice.  At each time step, each agent has a probability of $\gamma$, scaled by time step $\delta t$, of adding its own color graffiti at its current  location. At each time step, the graffiti also decays at a rate $\lambda>0$, similarly scaled by $\delta t$. Therefore, the amount of red graffiti at site $(x, y) \in S$ at time $t + \delta t$ is:
{\footnotesize
\begin{equation*}
g_A(x,y,t+\delta t)=g_A(x,y,t) - (\lambda \cdot \delta t ) g_A(x,y,t)  + (\gamma \cdot \delta t) n_A(x,y,t).
\end{equation*}
}
Again dividing by $l^2$ to convert $g_i(x,y,t)$ into density $\xi_i(x,y,t)$ for $i \in \{A,B\}$, we arrive at the graffiti density update rules:
{\footnotesize
\begin{align}
\xi_A(x,y,t+\delta t)=\xi_A(x,y,t) - (\lambda \cdot \delta t ) \xi_A(x,y,t)  + (\gamma \cdot \delta t) \rho_A(x,y,t)\\
\xi_B(x,y,t+\delta t)=\xi_B(x,y,t) - (\lambda \cdot \delta t ) \xi_B(x,y,t)  + (\gamma \cdot \delta t) \rho_B(x,y,t).
 \label{D:discrete_graffiti_A}
\end{align}
}

\subsection{\label{S:phases}Phases and an Order Parameter}  
When our model is simulated, we observe two possible states: a well-mixed state where agents and graffiti of both colors are uniformly distributed throughout the lattice, and a segregated state where red agents and red graffiti separate from blue agents and blue graffiti.  In this section, we explore these two possible phases for our model, defining an order parameter to quantify the distinction between the two states.

\subsubsection{\label{SS:expected_density}Expected Agent Density}
In the \emph{well-mixed state}, the agents and graffiti are roughly uniformly distributed on the lattice $S$, and each location $(x,y)$ is likely to have both red and blue agents and graffiti present. The movement of agents in a well-mixed state resembles an unbiased random walk. In the \emph{segregated state}, each gang clusters together to form territories, and the movement of agents becomes a biased random walk. In the segregated state, at each site $(x,y)$, there are usually agents of only one color present.

In a well-mixed state with $N_i$ agents from gang $i \in\{A,B\}$, the agents are uniformly distributed over the $L \times L$ lattice, and we can compute the expected agent density for gang $i \in \{A,B\}$ at any point $(x,y)$ on the lattice:
{\footnotesize
\begin{align}
E\left(\rho_i(x,y)\right)
&= \sum_{(x,y)\in S}  \rho_i(x,y) \times \frac{1}{L^2} \notag\\
   &= \sum_{(x, y) \in S}  \frac{n_i(x,y)}{l^2}\times l^2 \notag \\
   &= \sum_{(x,y) \in S} n_i(x,y) \notag\\
&= N_i. \label{E:mixed_energy_approximation}
\end{align}
}
In contrast, in the segregated state, the agents are separated into territories by color.  We make the assumption, numerically borne out in Sec. \ref{S:simulations}, that the agents are uniformly distributed within those territories. We consider all of the red territory as a sublattice denoted by $S_A \subset S$; the area of the sublattice $S_A$ is denoted by $R_A$. We note that the sublattice $S_A$ may not be connected. Similarly, we assume that blue territory considered all together is sublattice $S_B$ with area $R_B$.  We further assume that in a well-segregated state there are no empty sites, so that $S=S_A \cup S_B$.  These assumptions are reasonable from our numerical simulations; see Sec. \ref{S:simulations} for details. 

Under these assumptions, in the segregated state, the expected density of of red agents is:
\begin{equation*}
E\left(\rho_A(x,y)\right) = \sum_{(x,y)\in S}  \rho_A(x,y) \times \frac{1}{L^2}.
\end{equation*}
\noindent Splitting the lattice into sublattices $S_A$ and $S_B$ gives:
\begin{align*}
E\left(\rho_A(x,y)\right)&= \sum_{(x,y)\in S_A}  \frac{\rho_A(x,y)}{R_A} + \sum_{(x,y)\in S_B} \frac{\rho_A(x,y)}{R^B}\\
 &= \sum_{(x,y)\in S_A} \frac{\rho_A(x,y)}{R_A}\\
  &= \sum_{(x,y)\in S_A} \frac{n_A(x,y)}{l^2R_A}\\
&=  \frac{N_A}{l^2R_A} \text{, where } (x,y) \in S_A.
\end{align*}
\noindent The same argument holds for blue agents. Hence, for $i \in \{A,B\}$: 
\begin{equation}
E \left( \rho_i(x,y) \right) =
\begin{cases}
      \frac{N_i}{l^2R_i}, & (x,y) \in S_i \\
      0, & (x,y) \notin S_i.
\end{cases}
\label{E:segregated_energy_approximation_a}
\end{equation}
Finally, we note that in our model, the areas dominated by gangs $A$ and $B$ are nearly equal if we begin with $N_A = N_B$.  Thus, the assumption that $R_i = \frac{L^2}{2}$ is reasonable, giving:
{\footnotesize
\begin{equation}
E \left( \rho_i(x,y) \right) =
\begin{cases}
      2N_i,& (x,y) \in S_i \\
      0, & (x,y) \notin S_i.
\end{cases}
\label{E:segregated_energy_approximation_a_2}
\end{equation}
}

\subsubsection{An Order Parameter} 
To examine the phase transition, we define an order parameter at time $t$:
{\footnotesize
\begin{equation}
\mathcal{E}(t) =  \left(\frac{1}{2LN}\right)^2 \sum_{(x, y) \in S} \sum_{ (\tilde x,\tilde y) \sim ( x, y)} \left( \rho_A(x,y,t) - \rho_B(x,y,t) \right) \left( \rho_A(\tilde x, \tilde y, t) - \rho_B(\tilde x, \tilde y, t) \right).
\label{E:Energy_Equation}
\end{equation}
}
This order parameter is akin to a magnetization for the system, and is similar in form to the Hamiltonian for the Ising Model. The summand is positive if neighboring sites are dominated by the same color, and negative if they are dominated by the opposite colors.  The coefficient at the front normalizes the sum, so that the maximum value is $1$ independent of the lattice size and number of agents. In the segregated state, agents from each color cluster together to form territories, and at each site $(x,y)$ there is only one color present. This forces the term inside each of the sets of parentheses in equation \eqref{E:Energy_Equation} to be large in magnitude, with the sign in both cases very likely to be identical; once multiplied together, the result would be large and positive. However, in the well-mixed state, the agents of both gangs are uniformly distributed over all sites. This means that the terms inside the first and second brackets of equation \eqref{E:Energy_Equation} both tend to be very small, and the signs are unlikely to agree. Hence, once multiplied together, the result is very small in magnitude and, after summation, the order parameter is very close to zero. 

We now calculate an approximation of the order parameter for the well-mixed and segregated states. Similar to the expected value approximation and for the same reasons, we drop the time $t$ from the notation. Starting with the well-mixed case, the approximated order parameter is:
{\footnotesize
\begin{equation*}
\mathcal{E} = \left(\frac{1}{2LN}\right)^2 \sum_{(x, y) \in S} \sum_{ (\tilde x,\tilde y) \sim ( x, y)} \left(\rho_A(x,y) - \rho_B(x,y)\right) \left(\rho_A(\tilde x, \tilde y) - \rho_B(\tilde x, \tilde y) \right).
\end{equation*}
}
\noindent Using equation (\ref{E:mixed_energy_approximation}), the assumption that in a well-mixed state the distribution of agents is equal for all sites, and that each site has four neighbors,
{\footnotesize
\begin{align}
\mathcal{E} &= \left(\frac{1}{2LN}\right)^2 \sum_{(x, y) \in S} \left( N_A - N_B \right) \left( 4N_A - 4N_B \right) \notag \\
	  	  &= \left(\frac{1}{LN}\right)^2 \sum_{(x, y) \in S} \left( N_A - N_B \right)^2 \notag \\
	   &= \left(\frac{1}{LN}\right)^2 L^2 \left( N_A - N_B \right)^2 \notag \\
	  &= \frac{1}{N^2} \left( N_A - N_B \right)^2 =0,
\end{align}
}
where we have assumed in the last equality that $N_A = N_B$. Therefore, the order parameter in a well-mixed state is approximately zero.

The order parameter of the system in a completely segregated state can be derived similarly. Splitting the sum \eqref{E:Energy_Equation} over the regions  $S_A$ and  $S_B$, we see that:
{\footnotesize
\begin{align}
    \mathcal{E} &= \begin{aligned}[t]
   \left(\frac{1}{2LN}\right)^2 \Bigg[ \sum_{(x, y) \in S_A} \sum_{ (\tilde x,\tilde y) \sim ( x, y)} &\left(\rho_A(x,y) - \rho_B(x,y)\right)*\\
		&\left(\rho_A(\tilde x, \tilde y) - \rho_B(\tilde x, \tilde y) \right)
		\end{aligned} \notag \\
     &+ \begin{aligned}[t]
		\sum_{(x, y) \in S_B} \sum_{ (\tilde x,\tilde y) \sim ( x, y)} &\left(\rho_A(x,y) - \rho_B(x,y)\right)*\\
		&\left(\rho_A(\tilde x, \tilde y) - \rho_B(\tilde x, \tilde y) \right) \Bigg].
		\end{aligned} \notag
\end{align}
}
We substitute in the expectation of $\rho_A$ and $\rho_B$ for each sublattice; letting $R_i=|S_i|$ and ignoring the boundaries of the regions, we find:
{\footnotesize
\begin{align}
	\mathcal{E} &\approx \frac{1}{4}\left(\frac{1}{LN}\right)^2  \Bigg[\sum_{(x, y) \in S_A} \left( \frac{N_A}{l^2R_A} \right) \left(\frac{4N_A}{l^2R_A}\right) + \sum_{(x, y) \in S_B} \left(\frac{N_B}{l^2R_B}\right)\left(\frac{4N_B}{l^2R_B}\right)\Bigg] \notag \\
&=  \frac{1}{4}\left(\frac{l}{N}\right)^2 \frac{4}{l^4} \Bigg[\sum_{(x, y) \in S_A} \left( \frac{N_A}{R_A} \right)^2  + \sum_{(x, y) \in S_B} \left(\frac{N_B}{R_B}\right)^2 \Bigg] \notag \\
	&= \left(\frac{L}{N}\right)^2 \left[ R_A\left(\frac{N_A}{R_A}\right)^2 + R_B\left(\frac{N_B}{R_B}\right)^2\right] \notag \\
	&= \left(\frac{L}{N}\right)^2  \left(\frac{N_A^2}{R_A} + \frac{N_B^2}{R_B} \right). \notag
\end{align}
}
By assumption, we have perfect segregation with $N_A = N_B$, thus
{\footnotesize
\begin{align}
  \mathcal{E} 
	&\approx \left(\frac{L N_A}{N}\right)^2 \left( \frac{1}{R_A} + \frac{1}{R_B} \right) \notag \\
	&= \left(\frac{L}{2}\right)^2  \left( \frac{1}{R_A} + \frac{1}{R_B} \right). \notag
\end{align}
}
We substitute $R_A = R_B = \frac{L^2}{2}$ to find:
{\footnotesize
\begin{align}
	\mathcal{E} 
	&\approx 1.
\end{align}
}
Therefore, the order parameter parameter ranges from close to zero in a well-mixed state to close to one in a completely segregated state.

%%----------------------------------------------------------------
\section{\label{S:results}Results and Discussion}
\subsection{\label{S:simulations}Simulations of the Discrete Model} 

\subsubsection{Well-Mixed State}
\begin{figure}[!htp]
        \centering
        \subfigure{
                \includegraphics[width=0.30\linewidth,,keepaspectratio]{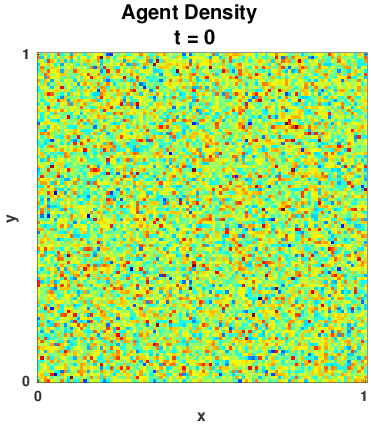}
                }
        \subfigure{
                \includegraphics[ width=0.30\linewidth,keepaspectratio]{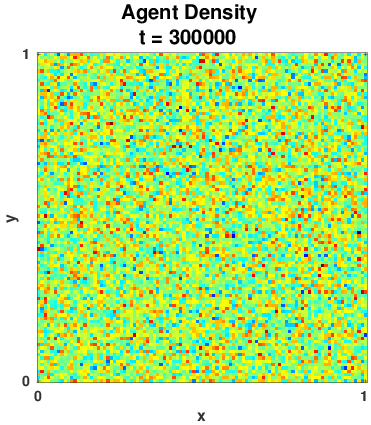}
                }\\
        \subfigure{
               \includegraphics[width=0.30\linewidth,,keepaspectratio]{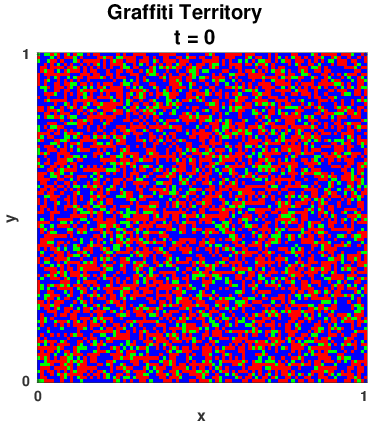}
               }
        \subfigure{
                \includegraphics[ width=0.30\linewidth,keepaspectratio]{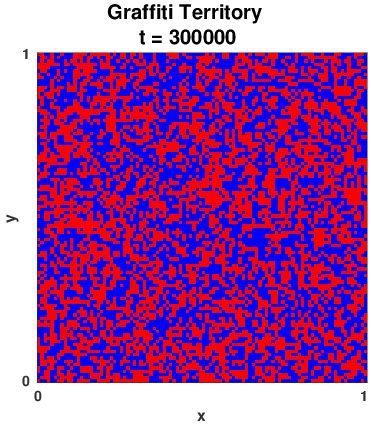}
                }
       \caption[Temporal evolution of the agent density on the left, and the temporal evolution for the graffiti territory on the right for a well-mixed state.]{Temporal evolution of the agent density on the left, and the temporal evolution for the graffiti territory on the right for a well-mixed state. Here we have  $N_A = N_B = 100,000$, with $\lambda = \gamma =0.5$, $\beta = 1 \times 10^{-6}$, $\delta t = 1$ and the lattice size is $100 \times 100$. It is clearly that the system remains well-mixed over time for these parameters, since we see neither red nor blue patches developing.}
\label{fig:64NotSegF1}
\end{figure}

We first consider the system in a well-mixed phase, with a small $\beta$ value.  We let $\beta = 1 \times 10^{-6}$, and evolve the system according to the discrete model; the resulting lattices are shown in Fig. \ref{fig:64NotSegF1}. The lattices in top row of Fig. \ref{fig:64NotSegF1} represent the agent density over time, while the bottom lattices represent the temporal evolution of the graffiti territory. The agent plots show how many agents of each gang are on the site; the higher the ratio of gang A to gang B, the more red the site appears, and the higher the ratio of gang B to gang A, the more blue it appears.  When the ratio is close to one, the site appears green. The graffiti territory plots show which gang has more graffiti on a particular site; if there is more graffiti from gang $A$ than gang $B$ on a site, then the site will be marked by the color red, and in the opposite situation, the site is marked by the color blue. We also assign the color green if there is exactly the same amount of graffiti present from both gangs at a site.

It is evident from Fig. \ref{fig:64NotSegF1} that we do not have segregation for $\beta = 1 \times 10^{-6}$, as we neither see red nor blue patches developing. In fact, the movement of each agent in this simulation resembles an unbiased random walk on a two-dimensional lattice. The agents' random walks are a direct result of the definition of the gang movement (\ref{D:probability_agent_moves_A}). If $\beta$  is sufficiently small, then the probability of an agent moving from $(x_1,y_1)$ to  $(x_2,y_2)$ approaches $\frac{1}{4}$, and therefore the movement approximates an unbiased random walk.  In effect, the parameter $\beta$ dampens agents' response to variations in the graffiti density $\xi$. Larger $\beta$ values amplify the variations of the graffiti density, leading to a phase transition. This phase transitions is studied in Sec. \ref{subsection:Energy}.

\begin{figure}[!htb]
        \centering
        \subfigure{
               \includegraphics[width=.45\linewidth,keepaspectratio]{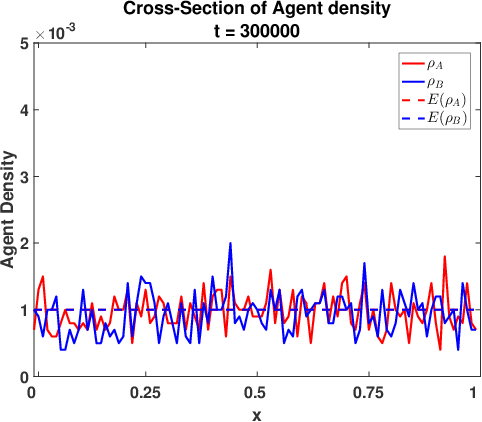}%{../Figures/NotSegF5k301.pdf}
               }
        \subfigure{
               \includegraphics[width=.45\linewidth,keepaspectratio]{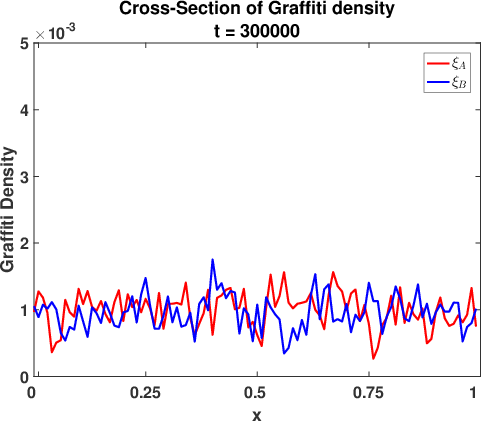}%{../Figures/NotSegF6k301.pdf}
               }
        \caption[Temporal evolution of a cross-sectional slice of the agent and graffiti densities for a well-mixed state.]{Temporal evolution of a cross-sectional slice of the agent and graffiti densities for a well-mixed state. Here we have  $N_A = N_B = 100,000$, with $\lambda = \gamma =0.5$, $\beta = 1 \times 10^{-6}$, $\delta t = 1$ and the lattice size is $100 \times 100$. The slice occurred at the $25^{th}$ row. It is clearly seen that the agents and the graffiti remain well-mixed over time for these parameters, and that our predicted agent expectation is a good approximation for the agent density.}
        \label{fig:64NotSegF5}
\end{figure}

Before investigating how larger $\beta$ values affect our system, let us examine the well-mixed phase more closely by taking a cross-sectional slice which shows the distribution of agent density on a lattice row. In these cross-sections, the effects of the stochasticity of the simulations can be seen quite clearly. The cross-sectional slices over time for $\beta = 1 \times 10^{-6}$ are shown in Fig. \ref{fig:64NotSegF5}; behind the cross-section for the agents, we also plot the expected agent density for the well-mixed phase approximated in equation \eqref{E:mixed_energy_approximation}. Note that we cannot consider an ensemble average for the cross-sectional slice because, due to the stochasticity in the model, each simulation could have a different territory evolution, leading to vastly diverging agent and graffiti distributions. Taking an ensemble average is therefore likely to result in the density distributions being uniformly distributed over time, obfuscating the territorial development and leading to incorrect conclusions.

From the cross-sectional slice in Fig. \ref{fig:64NotSegF5}, we notice that the agent and graffiti densities are roughly uniformly distributed across the row, with noise inherent to the stochastic agent-based simulations. It is clear from the figure that the agent density is quite close to $1 \times 10^5$, which is the expected agent density for a well-mixed phase.

\subsubsection{Well-Segregated State} \label{subsection:segregated}
\begin{figure}[!htb]
        \centering
        \subfigure{
               \includegraphics[width=.22\linewidth,keepaspectratio]{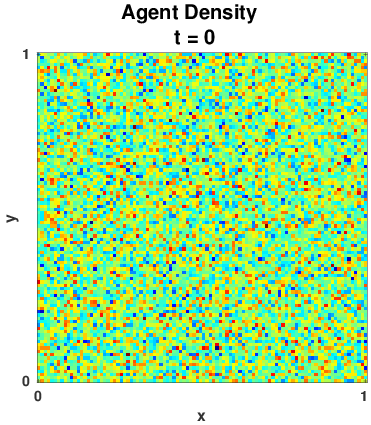}%{../Figures/Seg100F1k0.pdf}
               }
        \subfigure{
                \includegraphics[width=.22\linewidth,keepaspectratio]{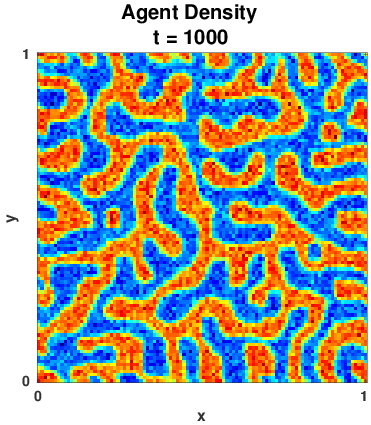}%{../Figures/Seg100F1k2.pdf}
                }
        \subfigure{
            \includegraphics[ width=.22\linewidth,keepaspectratio]{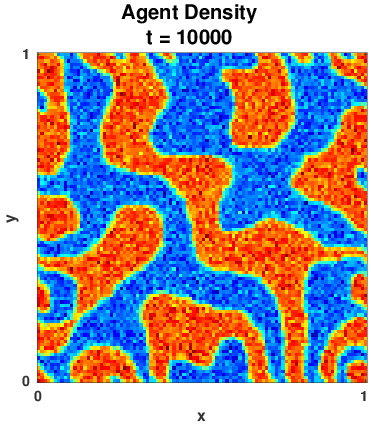}%{../Figures/Seg100F1k11.pdf}
            }
        \subfigure{
                \includegraphics[ width=.22\linewidth,keepaspectratio]{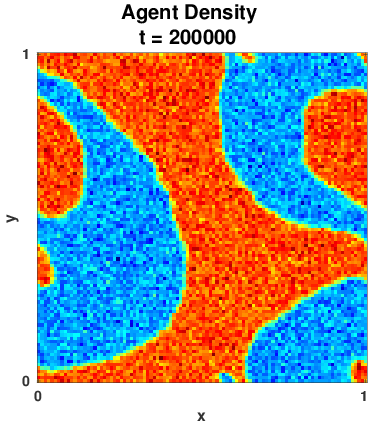}%{../Figures/Seg100F1k201.pdf}
				}
        \subfigure{
                \includegraphics[width=.22\linewidth,keepaspectratio]{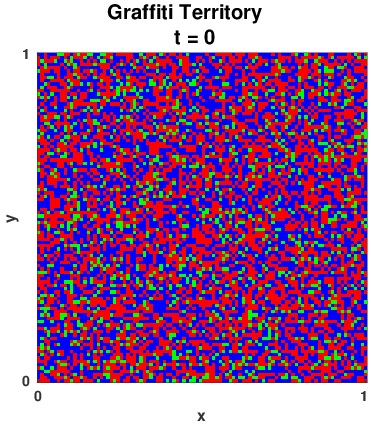}%{../Figures/Seg100F4k0.pdf}
                	}
        \subfigure{
               \includegraphics[width=.22\linewidth,keepaspectratio]{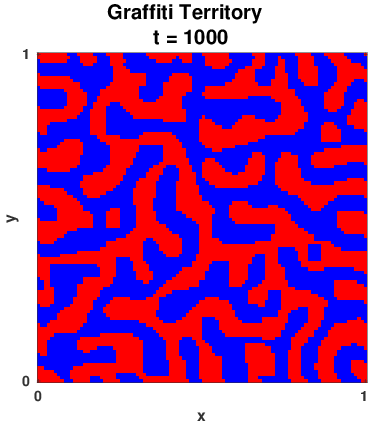}%{../Figures/Seg100F4k2.pdf}
               		}
        \subfigure{
                \includegraphics[ width=.22\linewidth,keepaspectratio]{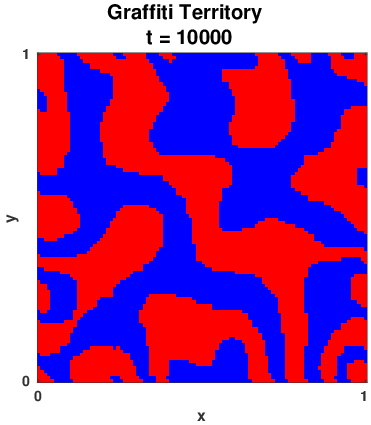}%{../Figures/Seg100F4k11.pdf}
                	}
        \subfigure{
               \includegraphics[ width=.22\linewidth,keepaspectratio]{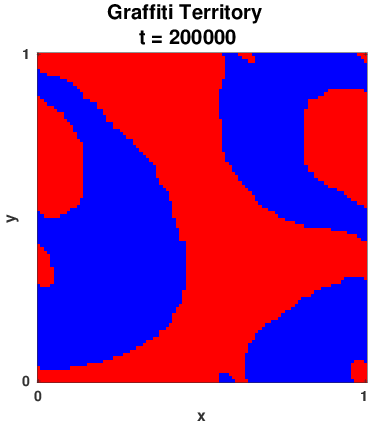}%{../Figures/Seg100F4k201.pdf}
               		}
        \caption[Temporal evolution of the agent density lattice and  the territory dominated by the gang's graffiti for a segregated state.]{Temporal evolution of the agent density lattice (top) and territory dominated by the gang's graffiti (bottom) for a segregated state. Here we have $N_A = N_B = 100,000$, with $\lambda = \gamma =0.5$, $\beta = 2 \times 10^{-5}$, $\delta t = 1$ and the lattice size is $100 \times 100$. It is clearly seen that both the agents and the territory dominated by graffiti segregate over time for these parameters.}
        \label{fig:64SegF1}
\end{figure}

We now increase the value of beta twenty-fold to $2 \times 10^{-5}$, while maintaining the lattice size and the number of agents as in the well-mixed state described above. Four time points of the simulations are shown in Fig. \ref{fig:64SegF1}. We observe that the initial state of the system is well-mixed; over time, we see that the red and blue agents start to cluster together and form all-red and all-blue areas, coarsening over time. Thus, $\beta = 2 \times 10^{-5}$ is large enough to produce a segregated phase, changing the agents' movement from an unbiased random walk to a biased random walk. This indicates that the critical $\beta$ or the phase transition should occur somewhere in the interval $\beta \in (1 \times 10^{-6},~ 2 \times 10^{-5})$.

\begin{figure}[!htb]
        \centering
        \subfigure{
                \includegraphics[width=.3\linewidth,keepaspectratio]{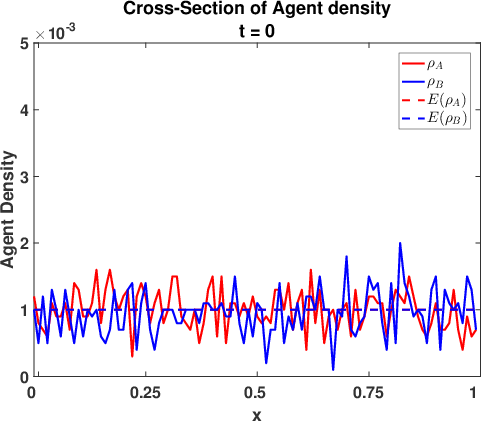}
                	}
        \subfigure{
                \includegraphics[width=.3\linewidth,keepaspectratio]{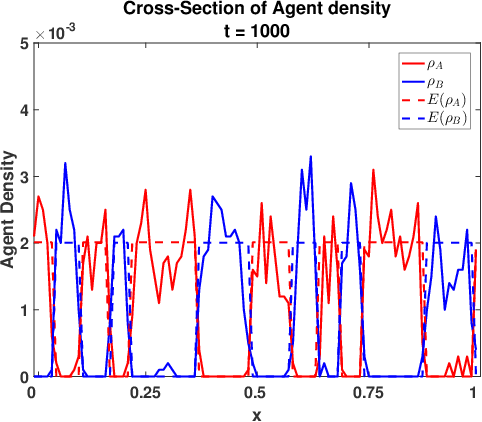}
                	}
        \subfigure{
                \includegraphics[width=.3\linewidth,keepaspectratio]{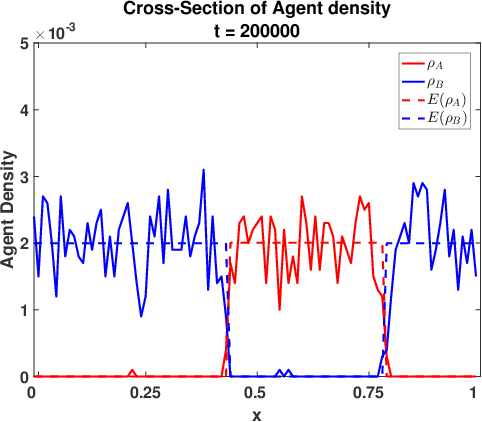}
                	}
                \subfigure{
                \includegraphics[width=.3\linewidth,keepaspectratio]{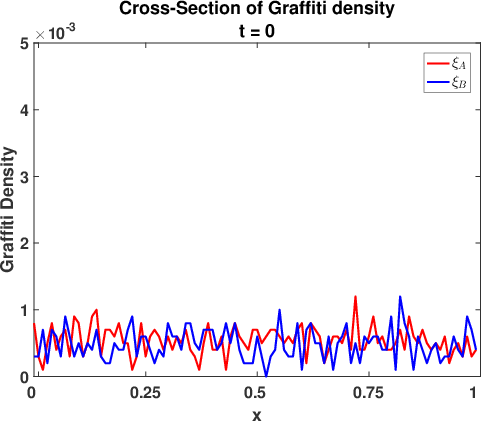}
                	}
        \subfigure{
                \includegraphics[width=.3\linewidth,keepaspectratio]{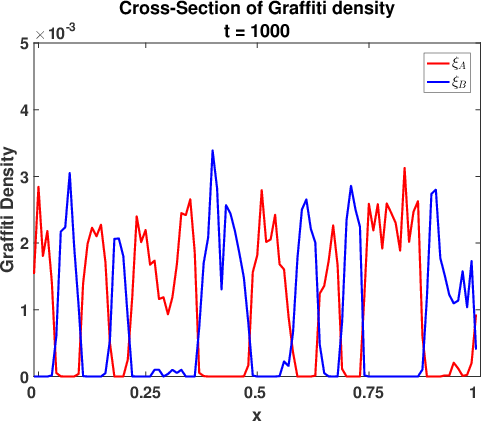}
                	}
                \subfigure{
                \includegraphics[width=.3\linewidth,keepaspectratio]{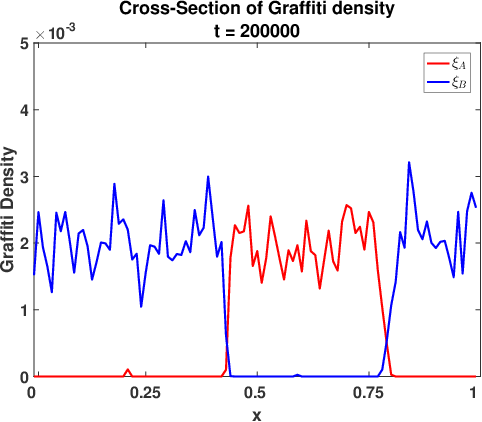}
                	}
        \caption[Temporal evolution of a cross-sectional slice of the agent graffiti densities for a segregated state.]   {Temporal evolution of a cross-sectional slice of the agent density (top) and  the graffiti density (bottom) for a segregated state. Here we have $N_A = N_B = 100,000$, with $\lambda = \gamma =0.5$, $\beta = 2 \times 10^{-5}$, $\delta t = 1$ and the lattice size is $100 \times 100$. Also, the slice occurred at the $25^{th}$ row. It is clearly seen that both the agents and graffiti segregate over time for these parameters and that our predicted agent expectation is good at approximating the agent density.}
        \label{fig:64SegF5}
\end{figure}

In Fig. \ref{fig:64SegF5}, we now consider cross-sectional slices for the segregated state in order to compare them against Fig. \ref{fig:64NotSegF5}. From equation \eqref{E:segregated_energy_approximation_a_2}, we know the expected agent densities for gang $i \in \{A,B\}$; these are plotted in the agent density plots as the red and blue dotted lines. Upon examining the cross-sectional slices in Fig. \ref{fig:64SegF5}, we observe that initially, the agents of both colors are uniformly distributed across the row. The uniform value of the agent density is roughly $1\times 10^5$ for both gangs. As time progresses, the agents begin to segregate, and increasingly large pockets of all-red or all-blue agents appear. Once segregated, we notice that the red agents and graffiti are nearly uniformly distributed in the all-red pockets with a value $2\times 10^5$, and vice versa for the blue agents and graffiti. The reason for the agent density doubling in value is that initially the gangs had to cover the entire row, but once segregated, the gangs had to cover only their own territory, which is approximately half of the row. This agrees with the expected agent density for a segregated state from equation (\ref{E:segregated_energy_approximation_a_2}); the roughness of the distributions is due to the stochastic nature of the simulation.

We end our discussion of the different states of the system by providing an illustrative example of how $\beta$ can amplify or dampen the effect of variations of the graffiti density $\xi$. Let us assume that the left, right, up and down neighbors of site $(x,y)$ have the following gang $B$ graffiti densities: \{$1 \times 10^5$, $0.55 \times 10^5$,  $0.5 \times 10^5$, $0.2 \times 10^5$\}. The graffiti density values used here are taken from the initial cross-sectional slice of the well-mixed phase in Fig. \ref{fig:64NotSegF5}. Using equation \eqref{D1:probability_agent_moves_A}, we check the probabilities of a red agent moving to a neighboring site for different $\beta$ values. We use the same $\beta$ values as the ones used in the well-mixed and well-segregated states, in addition to $\beta=6.5\times 10^{-6}$, which creates a partially segregated state. The resulting probabilities are summarized in Table \ref{table:beta_probability}. We notice that for the well-mixed $\beta$ value $1 \times 10^{-6}$, the probability of moving to any of the four neighboring sites is approximately $0.25$, although there is a large difference in the amount of graffiti on each of the sites. For a larger $\beta$ value, the probability of moving down is $0.4449$, which is much greater than the probability of moving to the left, which is $0.0898$. In this case, the graffiti differential causes the agents to undergo a biased random walk. This example clearly demonstrates how a larger $\beta$ amplifies the effects of the variations in the graffiti density, instigating the phase transition.

{\footnotesize
\begin{table}[!htb]
\centering
\begin{tabular}{ |c||c|c|c|c|}
\hline
\multirow{2}{*} {$\beta$} & $\xi_B = 1 \times 10^5$, & $ \xi_B  = 0.55 \times 10^5,$ & $\xi_B  = 0.5 \times 10^5,$ & $\xi_B  = 0.2 \times 10^5,$ \\
& $M_{\text{left}}$ & $M_{\text{right}}$ & $M_{\text{up}}$ & $M_{\text{down}}$ \\
\hline \hline
\multirow{2}{*}{$1 \times 10^{-6}$} & \multirow{2}{*}{0.2392} & \multirow{2}{*}{0.2502} & \multirow{2}{*}{0.2514} & \multirow{2}{*}{0.2591} \\ & & & & \\
\hline
\multirow{2}{*}{$6.5 \times 10^{-6}$} & \multirow{2}{*}{0.1849} & \multirow{2}{*}{0.2478}	& \multirow{2}{*}{0.2560} & \multirow{2}{*}{0.3111} \\ & & & & \\
\hline
\multirow{2}{*}{$2 \times 10^{-5}$} & \multirow{2}{*}{0.0898} & \multirow{2}{*}{0.2209} & \multirow{2}{*}{0.2442} & \multirow{2}{*}{0.4449} \\ & & & & \\
\hline
\end{tabular}
\caption[Probabilities of an agent from gang $A$ moving to a neighboring site for different $\beta$ values.]{Probabilities of an agent from gang $A$ moving to a neighboring site for different $\beta$ values. Clearly, a larger $\beta$ has a greater amplification effect on the variations in the graffiti density $\xi_B$.}
\label{table:beta_probability}
\end{table}
}

\subsubsection{Phase Transition in the Discrete Model} \label{subsection:Energy}

\begin{figure}[!htb]
        \centering
                \includegraphics[width=7cm,,keepaspectratio]{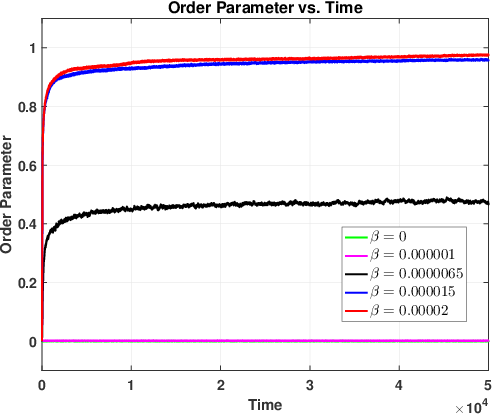}%{../Figures/f10.pdf}
        \caption[Temporal evolution of the order parameter  for different values of $\beta$.]{Temporal evolution of the order parameter  for different values of $\beta$. Here we have $N_A = N_B = 100,000$, with $\lambda = \gamma =0.5$, $\delta t = 1$ and the lattice size is $100 \times 100$. Clearly, for small values of $\beta$ the system was well-mixed and the order parameter stays close to zero through time. For sufficiently larger $\beta$ values, we have a segregated state and the order parameter  increases and levels off just below one.}
        \label{fig:10}
\end{figure}

We are now in a position to identify the critical $\beta$ at which the model undergoes a phase transition, using the order parameter defined in equation (\ref{E:Energy_Equation}). In Fig. \ref{fig:10}, we plot the ensemble average of four simulations of the order parameter over the course of a simulation for different values of $\beta$. We expect that system has a high order parameter value in a segregated state, and a low order parameter value in a well-mixed state.
In Fig. \ref{fig:10}, we see that for $\beta = 0$ and $\beta= 0.00001$, both of which correspond to the well-mixed state, the order parameter is almost zero over all time steps. For $\beta = 0.000015$ and  $\beta = 0.00002$, both of which correspond to the well-segregated state, we notice from the temporal evolution of the order parameter that the segregation occurs during the first $100,000$ time steps; afterwards the order parameter values equilibrate to around $0.95$ and remain there over time. Finally, for $\beta = 0.000065$, we observe from the order parameter that there is segregation, but it does not seem to be `complete'. This is reflected by the lower stabilized order parameter values for this case, which was approximately $0.5$.

\begin{figure}[!htp]
        \centering
        \subfigure{
                \includegraphics[width=.25\linewidth,keepaspectratio]{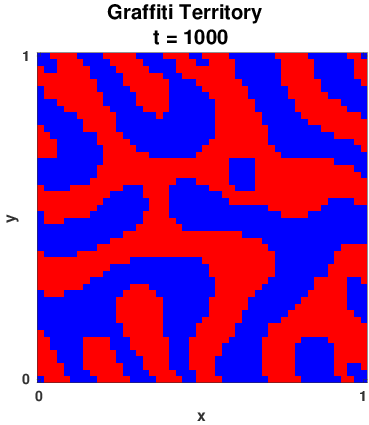}%{../Figures/Seg50F4k2.pdf}
					}
        \subfigure{
              \includegraphics[width=.25\linewidth,keepaspectratio]{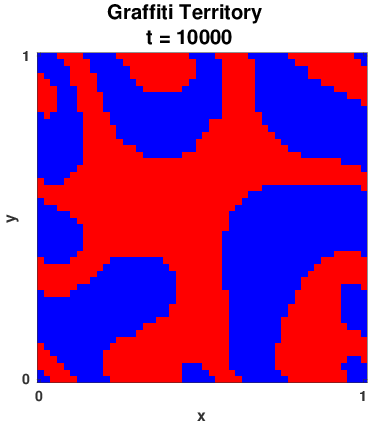}%{../Figures/Seg50F4k11.pdf}
              		}
        \subfigure{
             \includegraphics[ width=.25\linewidth,keepaspectratio]{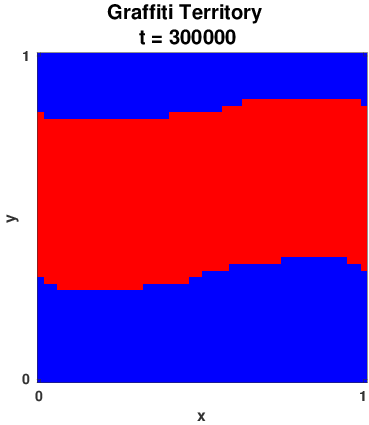}%{../Figures/Seg50F4k301.pdf}
             		}\\
        \subfigure{
             \includegraphics[width=.25\linewidth,keepaspectratio]{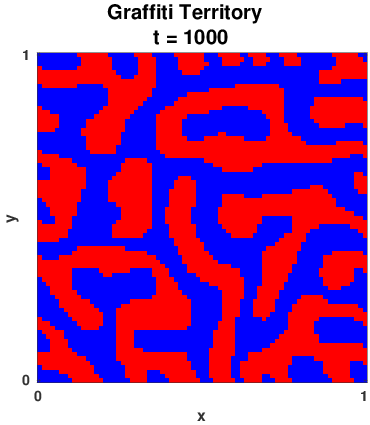}%{../Figures/Seg75F4k2.pdf}
             		}
       \subfigure{
               \includegraphics[ width=.25\linewidth,keepaspectratio]{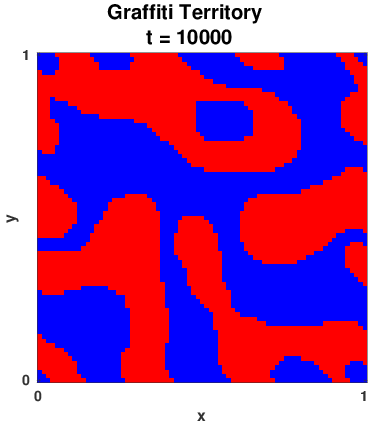}%{../Figures/Seg75F4k11.pdf}
               		}
        \subfigure{
            \includegraphics[ width=.25\linewidth,keepaspectratio]{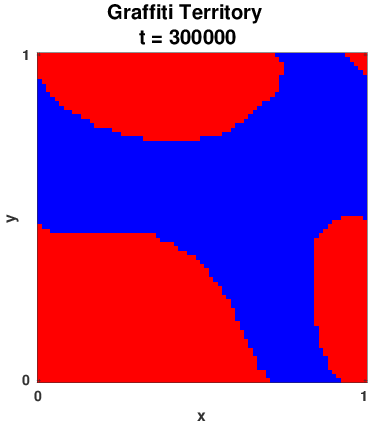}%{../Figures/Seg75F4k301.pdf}
            		}\\
        \subfigure{
           \includegraphics[width=.25\linewidth,keepaspectratio]{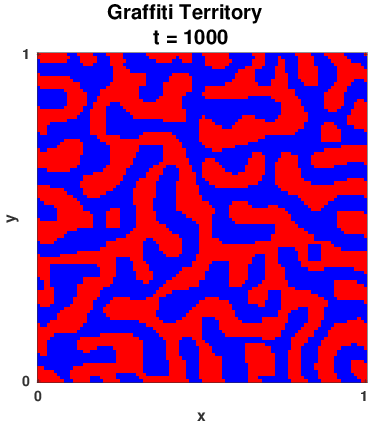}%{../Figures/Seg100F4k2.pdf}
           			}
        \subfigure{
              \includegraphics[ width=.25\linewidth,keepaspectratio]{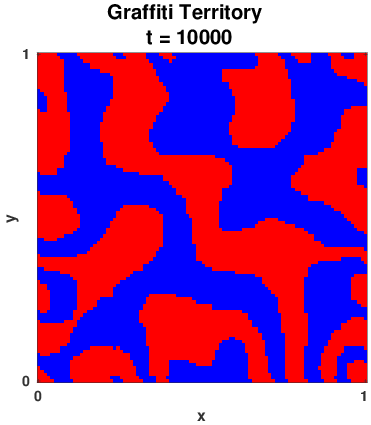}%{../Figures/Seg100F4k11.pdf}
              		}
        \subfigure{
              \includegraphics[ width=.25\linewidth,keepaspectratio]{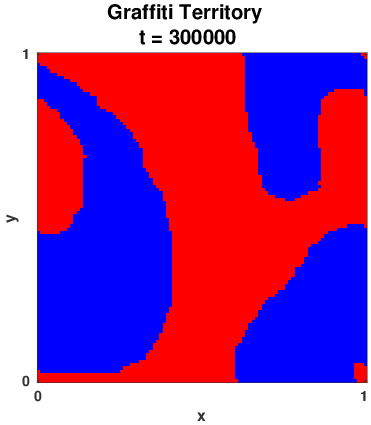}%{../Figures/Seg100F4k301.pdf}
              		}
               \caption[Temporal evolution of the territory dominated by the gang's graffiti for a segregated state  with different lattice sizes]{Temporal evolution of the territory dominated by the gang's graffiti for a segregated state  with different lattice sizes. Here we have $N_A = N_B = 100,000$, with $\lambda = \gamma =0.5$, $\beta = 2 \times 10^{-5}$ and $\delta t = 1$. The lattice sizes for the figure were $50 \times 50$ for the first row, $75 \times 75$ for the second row and $100 \times 100$ for the last row. It is clearly seen that the territory dominated by graffiti segregation is similar over time regardless of the lattice size.}
        \label{fig:64SameDensity}
\end{figure}

We now use the previously defined order parameter from equation (\ref{E:Energy_Equation}) to  find the critical value of $\beta$, $\beta_*$, which is the point at which the behavior of the system changes from well-mixed (disordered) to segregated (well-ordered). To locate $\beta_*$ numerically, we notice that the order parameter values of the system will eventually equilibrate to some constant (see Fig. \ref{fig:10}). We produce a phase transition plot by taking the final value of that order parameter constant and plotting it against different values of $\beta$. The critical parameter $\beta_*$ is the value at which the order parameter becomes nonzero; here, we approximate that value by taking $\beta_*$ be the point where the order parameter of the system surpasses $0.01$. Several phase transition plots for our system are shown in Fig. \ref{fig:f20andf22}.

The phase transition and $\beta_*$ might depend on the other parameters in the system, namely the total mass of the system ($N$), the graffiti production and decay rates ($\gamma$ and $\lambda$, respectively), the lattice dimension ($L$), and the time step ($\delta t$). To investigate how the phase transition depends on each of the parameters, we keep the other parameters fixed and vary only one.  In the first two figures of Fig. \ref{fig:f20andf22}, we can observe that the mass indeed affects the critical $\beta$. The left plot of the figure shows the phase transition for $N = 200,000$, with $N_A = N_B$, while the middle plot shows the phase transition for $N=100,000$, still with $N_A = N_B$.  We can observe that for the smaller mass, $\beta_*$ is almost double that of the simulations with the larger mass.  Hence, if the system has more agents, then the required $\beta$ for segregation is smaller. This makes sense in terms of our model, because if there are more agents in the system, then there will be more graffiti added, and each site will have a larger amount of graffiti. This implies that a smaller $\beta$ will be needed for the agents to react to that graffiti.

\subsubsection{The Role of Lattice Size and Time Step}

\begin{figure}
        \centering
            \subfigure{
            \includegraphics[width=0.3\linewidth,keepaspectratio]{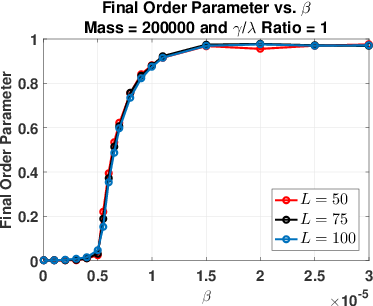}%{../Figures/f22.pdf}
            				}
            \subfigure{
            \includegraphics[width=0.3\linewidth,keepaspectratio]{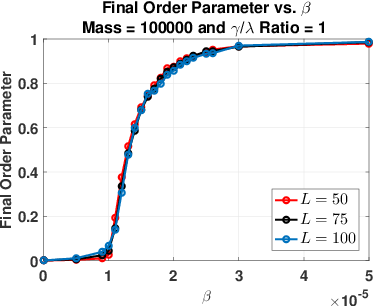}%{../Figures/f20.pdf}
            				}
            \subfigure{
            \includegraphics[width=0.28\linewidth,keepaspectratio]{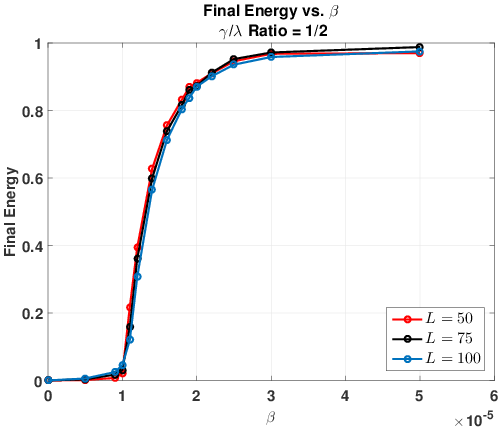}%{../Figures/f222.pdf}
            				}
               \caption[The order parameter  at the final time step against $\beta$ for different lattice sizes and number of agents.]{The order parameter  at the final time step against $\beta$ for different lattice sizes and number of agents. Here, we have $\lambda = \gamma =0.5$ with $\delta t = 1$. In the first figure, the number of agents $N_A = N_B = 100,000$, and in the last figure the number of agents $N_A = N_B = 50,000$. Clearly, the final order parameter for the different lattice sizes chosen is almost equal, and the critical $\beta$ is also equal for all three lattice sizes. Comparing the first and middle figures, we notice that the critical $\beta$ increased as the mass decreased. Comparing the first and bottom third, we notice that the critical $\beta$ increased as the ratio decreased.}
        \label{fig:f20andf22}
	\label{fig:RatioEnergyPlots}
\end{figure}

To examine how the lattice size can affect the phase transition of our system, we keep the number of agents, which is also referred as the system \emph{mass} $N$, fixed while we change the grid size. Note that in our system simulations the total area of the lattice always equals one, therefore by increasing the lattice size $L$, we make the grid finer and the sites smaller. For the visualization, we use the following lattice sizes: $50 \times 50$, $75 \times 75$, and $100 \times 100$. We also keep the system mass $N=200,000$ with $N_A=N_B$, and $\beta = 2 \times 10^{-5}$ fixed. The graffiti territory lattices for varying values of $L$ are shown in Fig. \ref{fig:64SameDensity}, while the order parameter evolution for the different lattice sizes can be seen in Fig. \ref{fig:f20andf22}. For the order parameter evolution, we plotted the ensemble average of four simulations.

From Fig. \ref{fig:f20andf22}, we observe that by keeping the system parameters fixed and only changing the grid size, the rate of segregation is unaffected, and that the coarsening rate does not depend on the grid size. This is important, because in Sec. \ref{S:derivation}, we will be deriving the continuum equations, and it is only natural to wonder how a finer grid may affect the discrete model. In Fig. \ref{fig:64SameDensity}, we see that a finer grid produces a `smoother' lattice visualization, as there are more sites, and they are smaller in size. Hence, the lattice for a finer grid is less `pixelated' and the territories have smoother boundaries.

\begin{figure}[htb!]
		\subfigure{
        \includegraphics[width=0.425\linewidth,keepaspectratio]{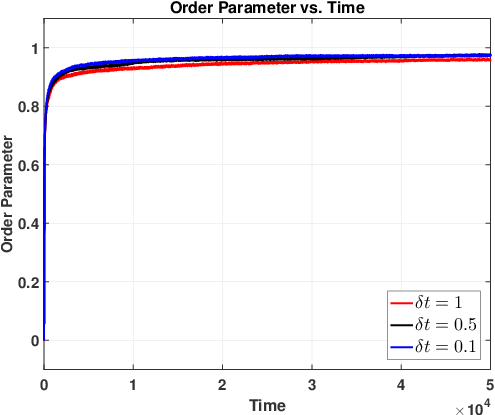}%{../Figures/energydt.pdf}
        			}
        \subfigure{			
        \includegraphics[width=0.425\linewidth,keepaspectratio]{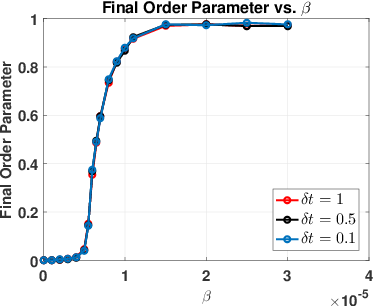}%{../Figures/f23.pdf}
        			}
        \caption[The phase transition for different time steps.]{The phase transition as the time step is changed.  In both figures, we have $N_A = N_B = 100,000$, with $\lambda = \gamma =0.5$ and the lattice size is $100 \times 100$. On the left, we see the temporal evolution of the order parameter for $\beta = 2 \times 10^{-5}$ (a segregated state). The evolution of the order parameter is observed to be the same for all three time steps shown. On the right, we see the order parameter at the final time step against $\beta$ for different time steps. It is clear to observe that the plots for the three different time steps chosen are almost identical, with the same critical $\beta_*$.}
\label{fig:f23}
\end{figure}

In Sec. \ref{S:derivation}, we will derive the continuum equations by taking the time step $\delta t$ to zero as we take the lattice spacing $l=\frac{1}{L}$ to zero. Hence, it is important to understand the role not only of the grid spacing but also of the time step in the dynamics of the discrete model.  To this end, we numerically observe the effects of a smaller time step in the discrete model. In Fig. \ref{fig:f23}, we visualize the temporal evolution of order parameter and the phase transition plot. We notice that decreasing the time step does not alter the critical $\beta$ for the phase transition or alter  the evolution of the order parameter, indicating that the size of the time step in our discrete model has little effect.

Comparing the first and last plots of Fig. \ref{fig:f20andf22}, we can see that the ratio $\frac{\gamma}{\lambda}$ also affects the critical $\beta$.  We see from the figure that as the ratio is halved from $1$ on the left to $\frac{1}{2}$ on the right, the shape of the phase transition is maintained but $\beta_*$ roughly doubles.  This, again, is unsurprising, since increasing the decay rate $\lambda$ in the ratio implies that the graffiti decays more quickly and there is less graffiti at each site, thereby forcing a higher $\beta$ value for segregation to occur. The same can be said about decreasing the graffiti rate $\gamma$.

We discuss the phase transitions again in Sec. \ref{chapter:stability}, where we examine how $\beta_*$ changes as we vary the parameters. There, we use the critical $\beta$ as a means to compare this discrete model with its continuum system counterpart, which is derived in the next section.

\subsection{\label{S:derivation}Derivation of the Continuum Model}
In order to better understand our system, in this section, we formally derive a system of corresponding continuum equations by taking the time step and the grid spacing to zero. 

\subsubsection{Continuum Graffiti Density}
The evolution equations for the graffiti density are easily found. Recalling the discrete model (\ref{D:discrete_graffiti_A}), and taking the limit $\delta t \rightarrow 0$, while assuming that the graffiti density $\xi_A$ is sufficiently smooth, the evolution equation for gang $A$'s graffiti becomes
\begin{equation}
\frac{\partial \xi_A}{\partial t}(x,y,t) = \gamma \rho_A(x,y,t) - \lambda \xi_A(x,y,t). \label{E:graffiti_continuum_eqn_1}
\end{equation}
The evolution equation for gang $B$'s graffiti follows identically.

\subsubsection{Continuum Agent Density}

Before deriving the continuum equation for the agent density, we define several quantities that will be of significant notational help. First, let us define $T_A$:
{\footnotesize
\begin{equation}
T_A(x,y,t) := \frac{e^{\beta \xi_B(x, y,t)}}{4 + l^2 \left( \left( \beta \nabla \xi_B( x, y,t) \right)^2 - \beta \Delta \xi_B( x,y,t) \right) }, \label{D:T_A}
\end{equation}
}
Hereafter, we will drop the $(x,y,t)$ as it is notationally superfluous. We will also need approximations to $\nabla T_A$ and $\Delta T_A$. Recalling the Taylor expansion
\begin{equation*}
\frac{1}{a+h} = \frac{1}{a} - \frac{h}{a^2} + \mathcal{O}(h^2).
\end{equation*}
and letting $a=4$ and $h=l^2 \left( \left( \beta \nabla \xi_B \right)^2 - \beta \Delta \xi_B \right)$, we can approximate $T_A$:
{\footnotesize
\begin{align}
T_A &= \frac{e^{\beta \xi_B}}{4}\left( 1 - \frac{l^2}{4}\left((\beta \nabla \xi_B)^2 - \beta \Delta \xi_B\right)  \right) + \mathcal{O}(l^4). \label{D:T_A_approximation}
\end{align}
}
Taking the gradient, we find:
{\footnotesize
\begin{align}
\nabla T_A &= \frac{e^{\beta \xi_B}}{4} \left( \beta \nabla \xi_B - \frac{l^2}{4}\left((\beta \nabla \xi_B)^3 + \beta^2 \nabla \xi_B \Delta \xi_B - \beta \nabla^3 \xi_B \right) \right) +\mathcal{O}(l^4). \label{D:T_A_gradient}
\end{align}
}
By taking the divergence, we find:
{\footnotesize
\begin{align}
\Delta T_A &= \nabla \cdot (\nabla T_A) \notag \\
&= \frac{e^{\beta \xi_B}}{4} \bigg( \left((\beta \nabla \xi_B)^2 + \beta \Delta \xi_B\right) - \frac{l^2}{4}\Big( 4 \beta^3 (\nabla \xi_B)^2 \Delta \xi_B  \notag \\
&+ \beta^2(\Delta \xi_B)^2  +(\beta \nabla \xi_B)^4 - \beta \nabla^4 \xi_B   \Big) \bigg)  + \mathcal{O}(l^4). \label{D:T_A_laplacian}
\end{align}
}
We derive $T_B, \nabla T_B$, and $\Delta T_B$ similarly.

We can now derive approximations to the probabilities of an agent arriving to the site $(x,y)$ from a neighboring site. Recall equation \eqref{D:probability_agent_moves_A},
\begin{equation}
M_A(\tilde x \rightarrow  x, \tilde y \rightarrow  y, t)  = \frac{e^{-\beta \xi_B(x, y,t)}}{\sum \limits_{(\tilde{\tilde x}, \tilde{\tilde y}) \sim (\tilde x,\tilde y)}e^{-\beta \xi_B(\tilde{\tilde x}, \tilde{\tilde y}, t)}},
\label{E:move}
\end{equation}
where $(\tilde{\tilde x}, \tilde{\tilde y})$ are the neighbors of site $(\tilde x,\tilde y)$. We use the discrete Laplacian to remove the influence of the neighbors' neighbors $(\tilde{\tilde x}, \tilde{\tilde y})$ in the denominator. Recalling the discrete Laplacian:
{\footnotesize
\begin{equation}
\sum \limits_{(\tilde{\tilde x}, \tilde{\tilde y}) \sim (\tilde x,\tilde y)}e^{-\beta \xi_B(\tilde {\tilde x},\tilde {\tilde y},t)} = 4e^{-\beta \xi_B(\tilde x,\tilde y,t)} + l^2 \Delta \left( e^{-\beta \xi_B(\tilde x,\tilde y,t)}\right) + \mathcal{O}(l^4),
\label{L:movement_from_neighbors_2}
\end{equation}
}
and noting that
{\footnotesize
\begin{align}
 \Delta e^{-\beta \xi_B(\tilde x,\tilde y,t)} &= \nabla \cdot \nabla \left(e^{-\beta \xi_B(\tilde x,\tilde y,t)}\right) \notag \\
&= \nabla \cdot \left( -\beta \nabla \xi_B(\tilde x,\tilde y,t) e^{-\beta \xi_B(\tilde x,\tilde y,t)}\right)  \notag \\
&= \left[ \left( \beta \nabla \xi_B(\tilde x,\tilde y,t) \right)^2 - \beta \Delta \xi_B(\tilde x,\tilde y,t) \right]e^{-\beta \xi_B(\tilde x,\tilde y,t)}, \label{L:movement_from_neighbors_3}
\end{align}
}
we combine (\ref{L:movement_from_neighbors_2}) with (\ref{L:movement_from_neighbors_3}) to give
{\footnotesize
\begin{align*}
&\sum \limits_{(\tilde{\tilde x}, \tilde{\tilde y}) \sim (\tilde x,\tilde y)} e^{-\beta \xi_B(\tilde {\tilde x},\tilde {\tilde y},t)} \\&= e^{-\beta \xi_B(\tilde x,\tilde y,t)} \left( 4 + l^2\left( \left( \beta \nabla \xi_B(\tilde x,\tilde y,t) \right)^2 - \beta \Delta \xi_B(\tilde x,\tilde y,t) \right) \right)+\mathcal{O}(l^4).
\end{align*}
}
Finally, we substitute it back into equation (\ref{E:move}), replacing the denominator to give
{\footnotesize
\begin{equation*}
\begin{split}
M_A(\tilde x \rightarrow  x, \tilde y \rightarrow  y, t) &= e^{-\beta \xi_B(x, y,t)} \left[ \frac{e^{\beta \xi_B(\tilde x,\tilde y,t)}}{4 + l^2 \left( \left( \beta \nabla \xi_B(\tilde x,\tilde y,t) \right)^2 -\beta \Delta \xi_B(\tilde x,\tilde y,t) \right) }  \right]  \\
&+\mathcal{O}(l^4).
\end{split}
\end{equation*}
}
However, we notice that the term in squared brackets takes the form of (\ref{D:T_A}), thus giving us the final form:
\begin{equation}\label{E:movement_from_neighbors}
M_A(\tilde x \rightarrow  x, \tilde y \rightarrow  y, t)=e^{-\beta \xi_B(x, y,t)} T_A(\tilde x,\tilde y,t)+\mathcal{O}(l^4).
\end{equation}
The result for gang B is similar, completing our notational toolbox.

We now formally derive the continuum equations for the agent density. Our main tools are the discrete Laplacian for approximating the influence of the neighbors of site $(x,y)$, and the approximations (\ref{E:movement_from_neighbors}). Recalling the discrete model (\ref{D:discrete_agents_A}) for the agent density, dividing both sides by $\delta t$, and noting that, at any time $t$, the movement probabilities away from a site sum to one gives us
{\footnotesize
\begin{align}
\frac{\rho_A(x,y,t + \delta t) - \rho_A(x,y,t)}{\delta t} = \frac{1}{\delta t} \Bigg[ &e^{-\beta \xi_B(x,y,t)} \sum_{ (\tilde x,\tilde y) \sim (x,y)}   \rho_A( \tilde x,  \tilde y, t)  T_A( \tilde x,  \tilde y, t)  \notag \\
 -& \rho_A(x, y, t) +\mathcal{O}(l^4) \Bigg].  \label{T:agents_continuum_eqn_1}
 \end{align}
}
We use the discrete Laplacian \eqref{L:movement_from_neighbors_2} to approximate the summation on the right, replacing the contribution from the neighboring sites with information from the current site:
{\footnotesize
\begin{align*}
\frac{1}{\delta t} \Bigg[ &e^{-\beta \xi_B(x,y,t)}\bigg(4 \rho_A(x,y,t)T_A(x,y,t)+l^2 \Delta \Big(\rho_A(x,y,t)T_A(x,y,t)\Big)\bigg)  \\
&-\rho_A(x,y,t) +\mathcal{O}(l^4) \Bigg].
\end{align*}
}
Now that the equation is governed entirely by quantities at site $(x,y)$ and time $t$, we can drop $(x,y,t)$ from the notation. Using definition (\ref{D:T_A}), we substitute the full expression for $T_A$ into the first term and simplify:
{\footnotesize
\begin{align}
&\frac{1}{\delta t} \Bigg[ 4\rho_A\left( \frac{1}{4 + l^2 \left( \left( \beta \nabla \xi_B \right)^2 - \beta \Delta \xi_B \right)} \right) - \rho_A+ l^2e^{-\beta \xi_B}\Delta \Big(\rho_A T_A\Big) +\mathcal{O}(l^4) \Bigg]. \label{T:agents_continuum_eqn_2}
\end{align}
}
Using a Taylor series expansion on the fractional term and substituting this back into expression (\ref{T:agents_continuum_eqn_2}) gives
{\footnotesize
\begin{equation}
\frac{1}{\delta t} \left[ 4\rho_A\left(\frac{1}{4} - \frac{l^2 \left( \left (\beta \nabla \xi_B\right) ^2 -\beta \Delta \xi_B \right) }{4^2} \right) - \rho_A  + l^2e^{-\beta \xi_B}\Delta \Big(\rho_A T_A\Big)  +\mathcal{O}(l^4)\right]. \notag
\end{equation}
}
Simplifying the expression yields
{\footnotesize
\begin{equation}
\begin{split}
\frac{\rho_A(x,y,t + \delta t) - \rho_A(x,y,t)}{\delta t} =& \frac{l^2}{\delta t} \left[ -\frac{\rho_A}{4}\left( \left( \beta \nabla \xi_B \right)^2 -\beta \Delta \xi_B \right)   + e^{-\beta \xi_B}\Delta \Big(\rho_A T_A\Big)  \right] \\
&+\mathcal{O}\left( \frac{l^4}{\delta t} \right).
 \label{T:agents_continuum_eqn_3}
 \end{split}
\end{equation}
}
We can further simplify by noting that
{\footnotesize
\begin{equation}
\Delta \Big(\rho_A T_A \Big) =   \Big(T_A \Delta \rho_A + 2 \nabla T_A \nabla \rho_A +\rho_A \Delta T_A\Big). \notag
\end{equation}
}
Therefore, substituting \eqref{D:T_A_approximation} through \eqref{D:T_A_laplacian} for $T_A$, $\nabla T_A$, and $\Delta T_A$, we have
{\footnotesize
\begin{align}
\Delta \Big(\rho_A T_A\Big) &= \frac{e^{\beta \xi_B}}{4} \Delta\rho_A  + \frac{2 \beta e^{\beta \xi_B} }{4} \nabla \xi_B  \nabla \rho_A +   \frac{e^{\beta \xi_B}}{4}\rho_A \left(\beta \Delta \xi_B +  (\beta \nabla \xi_B)^2 \right)  +\mathcal{O}(l^2)  \notag \\
&= \frac{e^{\beta \xi_B}}{4}  \left[\Delta \rho_A +  2 \beta \nabla \xi_B \nabla \rho_A  + \rho_A  \left( \left(\beta \nabla \xi_B\right)^2 + \beta \Delta \xi_B \right) \right] +\mathcal{O}(l^2). \label{T:agents_continuum_eqn_4}
\end{align}
}
\noindent Substituting (\ref{T:agents_continuum_eqn_4}) back into (\ref{T:agents_continuum_eqn_3}) gives us
{\footnotesize
\begin{align*}
\frac{\rho_A(x,y,t + \delta t) - \rho_A(x,y,t)}{\delta t} &=  \frac{l^2}{4 \delta t} \Bigg[ -\rho_A\left( (\beta \nabla \xi_B)^2 -\beta \Delta \xi_B \right) + \Delta \rho_A + 2 \beta \nabla \xi_B \nabla \rho_A  \\
  &\qquad  +  \rho_A  \left(\left(\beta \nabla \xi_B\right)^2 + \beta \Delta \xi_B\right) \Bigg] + \mathcal{O}\left(\frac{l^4}{\delta t}\right).
\end{align*}
}
Combining like terms, we find
{\footnotesize
\begin{align*}
\frac{\rho_A(x,y,t + \delta t) - \rho_A(x,y,t)}{\delta t}
&=\frac{l^2}{4\delta t}\Bigg[ \Delta \rho_A  + 2 \beta \nabla \cdot \Big(\rho_A \nabla \xi_B \Big) \Bigg] + \mathcal{O}\left(\frac{l^4}{\delta t}\right)   \\
&= \frac{l^2}{4\delta t} \nabla \cdot \Bigg[ \nabla \rho_A  + 2\beta \Big(\rho_A \nabla \xi_B \Big) \Bigg] +\mathcal{O}\left(\frac{l^4}{\delta t}\right).
\end{align*}
}
Assuming that the agent density $\rho_A$ is smooth, and that the limits
{\footnotesize
\begin{equation}
\begin{split}
l&\rightarrow 0, \\
\delta t &\rightarrow 0, \\
\frac{l^2}{\delta t} &\rightarrow D,
\end{split}
\end{equation}
}
\noindent hold, we arrive at the final form of the evolution equation for the density of red agents:
{\footnotesize
\begin{equation}
\frac{\partial \rho_A}{\partial t} =  \frac{D}{4} \nabla \cdot \Bigg[ \nabla \rho_A  + 2  \beta \Big(\rho_A \nabla \xi_B \Big) \Bigg]. \label{E:agents_continuum_eqn_1}
\end{equation}
}
An identical derivation holds for the density of blue agents.  Hence, our full system of continuum equations is
{\footnotesize
\begin{equation}\label{T:continuum_eqns}
\begin{cases}
\frac{\partial \xi_A}{\partial t}(x,y,t) = \gamma \rho_A(x,y,t) - \lambda \xi_A(x,y,t) \\
\frac{\partial \xi_B}{\partial t}(x,y,t) = \gamma \rho_B(x,y,t) - \lambda \xi_B(x,y,t) \\
\frac{\partial \rho_A}{\partial t}(x,y,t) =  \frac{D}{4} \nabla \cdot \Big[ \nabla \rho_A(x,y,t)  + 2  \beta \big(\rho_A(x,y,t) \nabla \xi_B(x,y,t) \big) \Big] \\
\frac{\partial \rho_B}{\partial t}(x,y,t) =  \frac{D}{4} \nabla \cdot \Big[ \nabla \rho_B(x,y,t)  + 2  \beta \big(\rho_B(x,y,t) \nabla \xi_A(x,y,t) \big) \Big],
\end{cases}
\end{equation}
}
\noindent with periodic boundary conditions. 

The dimensionless form of the continuum system is
{\footnotesize
\begin{equation}\label{E:NonDimensionalContinuumEquations}
\begin{cases}
\frac{\partial \xi_A}{\partial t} =  c  \rho_A -  \xi_A  \\
\frac{\partial \xi_B}{\partial t} =  c  \rho_B -  \xi_B  \\
\frac{\partial \rho_A}{\partial t} =  \frac{1}{4} \nabla_{ X} \cdot \Big[  \nabla_{ X}  \rho_A  + 2  \beta  \big( \rho_A \nabla_{ X}  \xi_B \big) \Big] \\
\frac{\partial  \rho_B}{\partial \tilde t} =  \frac{1}{4} \nabla_{ X} \cdot \Big[ \nabla_{ X}  \rho_B  + 2  \beta \big( \rho_B \nabla_{ X}  \xi_A \big) \Big].
\end{cases}
\end{equation}
}
Details of the nondimensionalization can be found in Appendix \ref{section:dimension}.

\subsection{\label{S:analyzing}Studying the Continuum Model} 
Now that a continuum version of the model has been derived, we have more tools with which to understand the model.  We first verify that the continuum system and the discrete system share the same uniform and segregated equilibrium solutions; then we perform a linear stability analysis around the uniform equilibrium solution to gain insight into the phase transition. 
\subsubsection{Steady-State Solutions} \label{subsection:steadystatesolutions}
Identifying the steady-state for the graffiti is straightforward: setting $\frac{\partial \xi_i}{\partial t}=0$ yields 
\begin{equation}
\xi_i = \frac{\gamma}{\lambda} \rho_i \mbox{ for } i \in \{A,B\}.
\label{E:stationaryGraffiti}
\end{equation}
\noindent Looking more closely at the equations for agent density, we note that steady-state solutions for the red gang must satisfy
\begin{align}
\nabla \rho_A(x,y,t)  + 2 \beta \Big(\rho_A(x,y,t) \nabla \xi_B(x,y,t) \Big) &= c_A \notag \\
\intertext{for $c_A \in R$. Using the equilibrium graffiti density \eqref{E:stationaryGraffiti}, we see that}
\nabla \rho_A(x,y,t)  + \frac{2 \beta \gamma}{\lambda} \Big(\rho_A(x,y,t) \nabla \rho_B(x,y,t) \Big) &= c_A. \label{E:steadystate_rho1}
\end{align}
and similarly for the blue gang. Thus, any form of $\rho_A(x,y,t)$ and $\rho_B(x,y,t)$ satisfying
\begin{equation}\label{T:continuum_eqns}
\begin{cases}
\xi_A(x,y,t) = \frac{\gamma}{\lambda} \rho_A(x,y,t)\\
\xi_B(x,y,t) = \frac{\gamma}{\lambda} \rho_B(x,y,t)\\
\nabla \rho_A(x,y,t)  + 2  \beta \big(\rho_A(x,y,t) \nabla \xi_B(x,y,t) \big) = c_A\\
\nabla \rho_B(x,y,t)  + 2  \beta \big(\rho_B(x,y,t) \nabla \xi_A(x,y,t) \big) = c_B,
\end{cases}
\end{equation}
\noindent is a steady-state solution of our system.

The well-mixed state, with all agent and graffiti densities uniformly distributed so that $\xi_i = \frac{\gamma}{\lambda} \rho_i$, is clearly a steady-state for our system.  Another obvious steady-state solution takes the following form:
{\footnotesize
\begin{equation} \label{E:steadystateequation}
\begin{split}
\xi_A &= \frac{\gamma}{\lambda} \rho_A \\
\xi_B &= \frac{\gamma}{\lambda} \rho_B \\
\rho_A &=  \begin{cases}
      c_A, & 0 < x < 0.5 \\
     0,~ &  0.5 < x < 1
\end{cases} \\
\rho_B &=  \begin{cases}
      0, & 0< x < 0.5 \\
     c_B,~ &  0.5 < x < 1.
\end{cases}
\end{split}
\end{equation}
}

\begin{figure}[!htp]
        \centering
                \subfigure{
                \includegraphics[width=.45\linewidth,keepaspectratio]{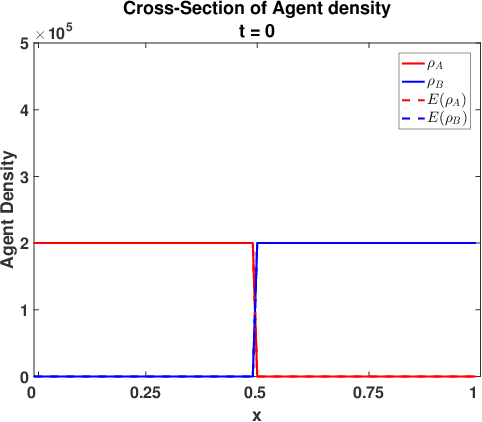}%{../Figures/dividedF5k0.pdf}
                				}
        \subfigure{
               \includegraphics[width=.45\linewidth,keepaspectratio]{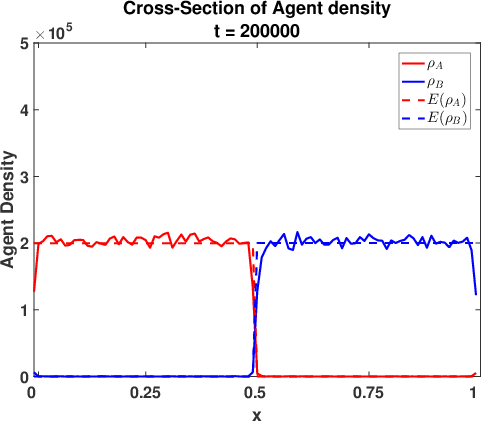}%{../Figures/dividedF5k4.pdf}
               		}
        \caption[Temporal evolution of the agent density lattice for a steady-state solution.]{Temporal evolution of an ensemble average of cross-sectional slices of the agent density for a steady-state solution. Here we have $N_A = N_B = 100,000$, with $\lambda = \gamma =0.5$, $\beta = 2 \times 10^{-5}$, $\delta t = 1$ and the lattice size is $100 \times 100$. We see that if the system started in its steady state, that is, the agents are initially segregated, then the system would remain in that state over time. The ensemble average of $40$ simulations was used. We clearly see that if the system starts in a steady-state solution, then it remains in that state over time.}
        \label{fig:steadystatedivided}
\end{figure}

The steady-state solution \eqref{E:steadystateequation} can be used to compare the discrete model and the continuum model. Starting our simulations with the agents completely segregated, we visualize the temporal evolution of the agent densities over time in Fig. \ref{fig:steadystatedivided}. This figure shows an ensemble average of $40$ cross-sectional slices for agent density, where we can see that the agent density remains constant over time and the density clearly follows the form of the steady-state solutions of equation \eqref{E:steadystateequation}. Note that in this context, unlike in Sec. \ref{subsection:segregated}, taking an ensemble average is sensible since we know the expected the solution. In the figure, we also notice that the graffiti is proportional to the agent density: $\xi_{A,B} = \frac{\gamma}{\lambda} \rho_{A,B}$.

\begin{figure}[!htp]
        \subfigure{
                \includegraphics[width=.4\linewidth,keepaspectratio]{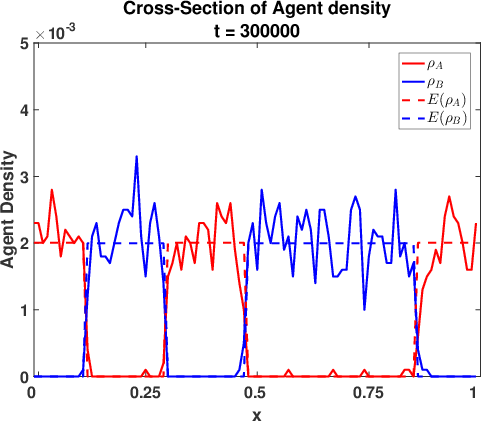}%{../Figures/Seg100RatioHalfF5k301.pdf}
                }
        \subfigure{
               \includegraphics[width=.4\linewidth,keepaspectratio]{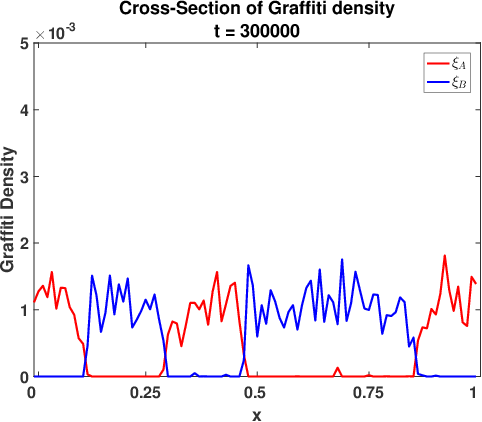}%{../Figures/Seg100RatioHalfF6k301.pdf}
				}
        \subfigure{
                \includegraphics[width=.4\linewidth,keepaspectratio]{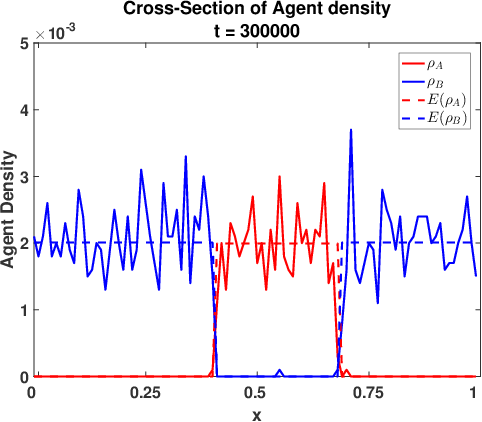}%{../Figures/Seg100F5k301.pdf}
                	}
        \subfigure{
               \includegraphics[width=.4\linewidth,keepaspectratio]{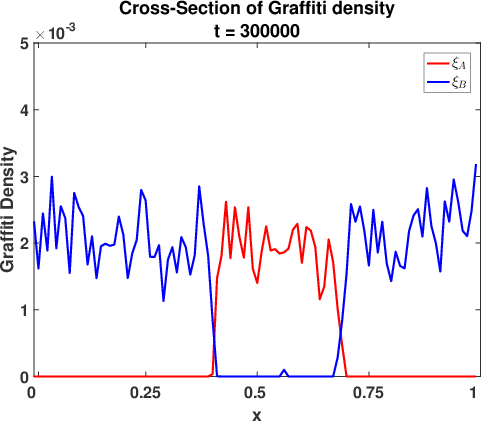}%{../Figures/Seg100F6k301.pdf}
					}
        \subfigure{
                \includegraphics[width=.42\linewidth,keepaspectratio]{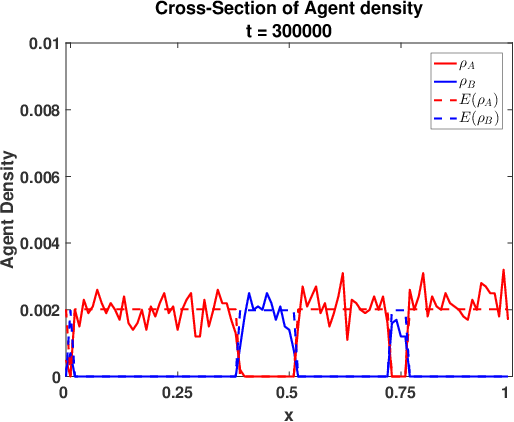}%{../Figures/Seg100Ratio2F5k301.pdf}
					}
        \subfigure{
               \includegraphics[width=.42\linewidth,keepaspectratio]{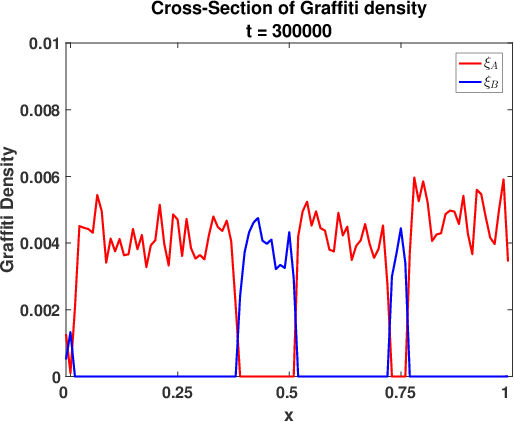}%{../Figures/Seg100Ratio2F6k301.pdf}
               		}
        \caption[The final time step of the agent and graffiti densities for a segregated state using different $\frac{\gamma}{\lambda}$ ratios.]{The final time step of the agent and graffiti densities for a segregated state using different $\frac{\gamma}{\lambda}$ ratios. Here we have $N_A = N_B = 100,000$, with $\beta = 2 \times 10^{-5}$, $\delta t = 1$ and the lattice size is $100$. The ratios for the figure were $\gamma = 0.25$ and $\lambda = 0.5$ for the first row, $\gamma = \lambda = 0.5$ for the second row and $\gamma = 0.5, \lambda = 0.25$ for the last row. Looking at the cross-sectional slices, we see that $\xi_A \approx \frac{1}{2} \rho_A$, $\xi_A \approx \rho_A$ and $\xi_A \approx 2\rho_A$ for the top, middle and bottom rows, respectively. Hence, the discrete model agrees with the steady-state solution of equation, $\xi_A = \frac{\gamma}{\lambda} \rho_A$.}
        \label{fig:steadystate}
\end{figure}

To test whether \eqref{E:stationaryGraffiti} holds more generally for the segregated state in the  discrete model, we simulate our system with different values for the ratio $\frac{\gamma}{\lambda}$, then comparing the resulting densities with that of equation \eqref{E:steadystateequation}. We use the same starting conditions as in Sec. \ref{S:discrete}, and at the final time step we take a cross-sectional slice over the entire lattice. These cross-sectional slices are visualised in Fig. \ref{fig:steadystate} for the following ratio values: $\frac{1}{2}$, $1$, and $2$. The figure indicates that the relationship \eqref{E:stationaryGraffiti} derived from the continuum equations holds generally for simulations of the segregated state in the discrete model.

\subsubsection{Linear Stability Analysis}\label{chapter:stability}
To help us better understand our system, we perform a linear stability analysis on the uniformly distributed equilibrium solution corresponding to the well-mixed state.  Taking the same approach as \cite{BLP2002,JBC2010,SDPTBBC2008,WLM1996}, we consider perturbations of the form $\epsilon = \delta e^{\alpha t}e^{ikx}$, with $\delta \ll 1$, so that our solution takes the form:
\begin{align}
\begin{cases}
\xi_A = \bar{\xi_A} + \delta_{\xi_A} e^{\alpha t}e^{ikx}\\%\label{E:dimensionalpertubation1}\\
\xi_B = \bar{\xi_B} + \delta_{\xi_B} e^{\alpha t}e^{ikx}\\%\label{E:dimensionalpertubation2}\\
\rho_A = \bar{\rho_A} + \delta_{\rho_A} e^{\alpha t}e^{ikx}\\%\label{E:dimensionalpertubation3}\\
\rho_B = \bar{\rho_B} + \delta_{\rho_B} e^{\alpha t}e^{ikx}.%\label{E:dimensionalpertubation4}
\end{cases}
\label{E:dimensionalperturbation}
\end{align}
For the well-mixed solution to be stable, $\alpha$ must be negative, forcing the perturbations to decay over time.

Substituting the perturbed steady-state \eqref{E:dimensionalperturbation} into the graffiti equation yields
{\footnotesize
\begin{align*}
\frac{\partial}{\partial  t} \left( \bar{\xi_A} + \delta_{\xi_A} e^{\alpha t}e^{ikx} \right) &= \gamma (\bar{\rho_A} +\delta_{\rho_A} e^{\alpha t}e^{ikx}) -\lambda (\bar{\xi_A} +\delta_{\xi_A} e^{\alpha t}e^{ikx} ).
\end{align*}
}
Since $\bar{\xi_A}$ is an equilibrium solution, $\frac{\partial  \bar{\xi_A}}{ \partial  t} = \gamma  \bar{\rho_A} - \lambda \bar{\xi_A} = 0$.  
Hence,
\begin{align}
\alpha \delta_{\xi_i} &= (\gamma \delta_{\rho_i}-\lambda \delta_{\xi_i}) \quad \mbox{ for } i \in \{A,B\}.\label{E:dimensionalstable1}
\end{align}

Substituting\eqref{E:dimensionalperturbation} 
into the evolution equation for the agent density gives us
{\footnotesize
\begin{align}
\frac{\partial}{\partial  t}  \left( \bar{\rho_A} + \delta_{\rho_A} e^{\alpha t}e^{ikx} \right) &=  \frac{D}{4} \Delta \left( \bar{\rho_A} + \delta_{\rho_A} e^{\alpha t}e^{ikx} \right) \notag \\
&\quad+  \frac{D\beta}{2} \nabla\cdot \Big( (\bar{\rho_A} + \delta_{\rho_A} e^{\alpha t}e^{ikx} ) \nabla (\bar{\xi_B} + \delta_{\xi_B} e^{\alpha t}e^{ikx}) \Big). \notag
\end{align}
\noindent Since our equilibrium solution is constant in both time and space,
\begin{align}
\alpha \delta_{\rho_A} e^{\alpha t}e^{ikx} &= \frac{-D|k|^2}{4} \delta_{\rho_A} e^{\alpha t}e^{ikx}  + \frac{D\beta}{2} \frac{d}{d x} \left( (\bar{\rho_A} + \delta_{\rho_A} e^{\alpha t}e^{ikx} ) (ik \delta_{\xi_B} e^{\alpha t}e^{ikx} )\right) \notag \\
&= \frac{-D|k|^2}{4} \left( \delta_{\rho_A} + 2 \beta \bar{\rho_A} \delta_{\xi_B} \right) e^{\alpha t}e^{ikx} +\mathcal{O}(\delta_{\rho_A}\delta_{\xi_B}). \notag
\end{align}
}
Neglecting the term  $\mathcal{O}(\delta_{\rho_A}\delta_{\xi_B})$,
\begin{equation}
\alpha \delta_{\rho_A} = \frac{-D|k|^2}{4} \left( \delta_{\rho_A} + 2 \beta \bar{\rho_A} \delta_{\xi_B} \right), \label{E:dimensionalstable3}
\end{equation}
and similarly for $\delta_{\rho_B}$.

Writing the linearized equations \eqref{E:dimensionalstable1} through \eqref{E:dimensionalstable3} in systems form,
\begin{equation*}
\begin{bmatrix}
-\lambda & 0 & \gamma & 0\\
0 & -\lambda & 0 & \gamma\\
0 & \frac{-\beta D\bar{\rho_A} |k|^2}{2} & \frac{-D|k|^2}{4} & 0\\
\frac{-\beta D \bar{\rho_B} |k|^2}{2} & 0 & 0 & \frac{-D|k|^2}{4} \\
\end{bmatrix} \begin{bmatrix}  \delta_{\xi_A}\\  \delta_{\xi_B}\\  \delta_{\rho_A}\\  \delta_{\rho_B} \end{bmatrix} = \alpha\begin{bmatrix}  \delta_{\xi_A}\\  \delta_{\xi_B}\\  \delta_{\rho_A}\\  \delta_{\rho_B} \end{bmatrix}.
\end{equation*}
gives us an equation of the form $\left(M - \alpha I_4 \right)\vec{\delta} = 0.$
This is an eigenvalue equation for the matrix $M$, and for it to have non trivial solutions (i.e. solutions where $\vec{\delta} \neq 0$), the determinant of $(M - \alpha I_4)$ must be zero. Thus,
\begin{equation*}
\begin{vmatrix}
-(\lambda+\alpha) & 0 & \gamma & 0\\
0 & -(\lambda+\alpha) & 0 & \gamma\\
0 & \frac{-\beta D \bar{\rho_A} |k|^2}{2} & -\left(\frac{D|k|^2}{4}+\alpha\right) & 0\\
\frac{-\beta D \bar{\rho_B} |k|^2}{2} & 0 & 0 & -\left(\frac{D|k|^2}{4}+\alpha\right) \\
\end{vmatrix}=0,
\end{equation*}
giving us the characteristic polynomial
{\footnotesize
\begin{equation*}
\begin{split}
f(\alpha) &= \frac{1}{16} \bigg( \left( (\lambda + \alpha)^2 - 4\beta^2\gamma^2 \bar{\rho_A}\bar{\rho_B}\right)D^2|k|^4 +8\alpha(\lambda + \alpha)^2(2\alpha +D|k|^2) \bigg) = 0.
\end{split}
\end{equation*}
}
Solving the characteristic polynomial gives us the following four eigenvalues:
{\footnotesize
\begin{numcases}{}
\alpha_{1,2} = -\frac{1}{8} \left( 4\lambda + D|k|^2 \pm \sqrt{16\lambda^2 - 8D(\lambda+4\beta \gamma \sqrt{\bar{\rho_A}\bar{\rho_B}})|k|^2 + D^2|k|^4}\right)  \label{E:dimensionalev1} \\
\text{and} \notag \\
\alpha_{3,4} = -\frac{1}{8} \left( 4\lambda + D|k|^2 \pm \sqrt{16\lambda^2 - 8D(\lambda-4\beta \gamma \sqrt{\bar{\rho_A}\bar{\rho_B}})|k|^2 + D^2|k|^4}\right). \label{E:dimensionalev2}
\end{numcases}
}

\begin{figure}[!htp]
        \centering
        \subfigure{
        \includegraphics[width=.45\linewidth,keepaspectratio]{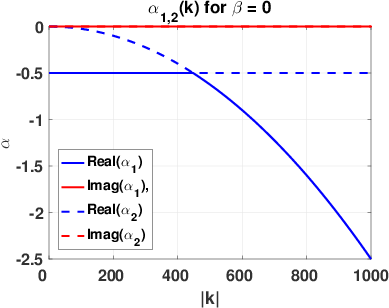}%{../Figures/lsa111.pdf}
        			}
         \subfigure{
        \includegraphics[width=.45\linewidth,keepaspectratio]{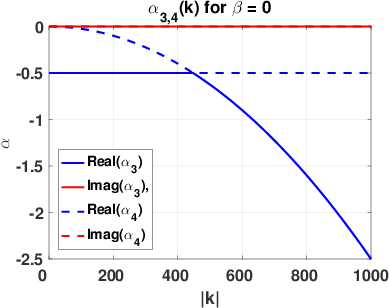}%{../Figures/lsa222.pdf}
					}
        \subfigure{
        \includegraphics[width=.45\linewidth,keepaspectratio]{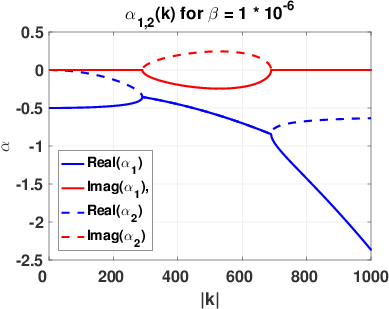}%{../Figures/lsa333.pdf}
					}
        \subfigure{
        \includegraphics[width=.45\linewidth,keepaspectratio]{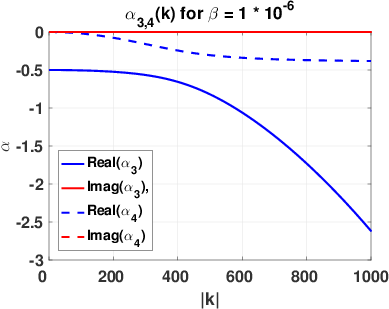}%{../Figures/lsa444.pdf}
        			}
        \subfigure{
        \includegraphics[width=.45\linewidth,keepaspectratio]{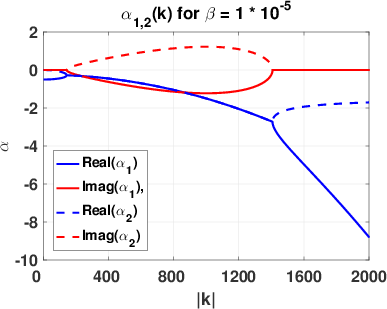}%{../Figures/lsa555.pdf}
        			}
        \subfigure{
        \includegraphics[width=.45\linewidth,keepaspectratio]{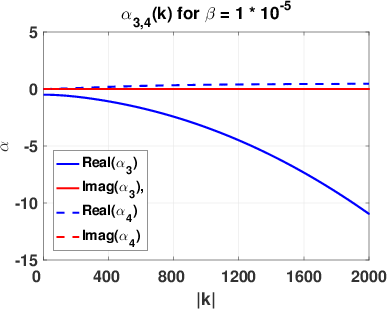}%{../Figures/lsa666.pdf}
        			}
       \caption[Eigenvalues for different values of $\beta$, plotted against wave number $k$.]{Eigenvalues for different values of $\beta$, plotted against wave number $k$. Here we have $D= 1 \times 10^{-4}$, the ratio $\frac{\gamma}{\lambda} = 1$, and the density $\rho_A =\rho_B = 100,000$.}
        \label{fig:dimensionalstability1}
\end{figure}

Fig. \ref{fig:dimensionalstability1} shows the four eigenvalues as functions of the wave number $k$. In the figure, we set $\frac{\gamma}{\lambda}=1$ and densities $\rho_A = \rho_B = 100,000$, which are the values that were used in the discrete simulations in Sec. \ref{S:simulations}.

To classify the linear stability of our equilibrium solutions, we determine when the eigenvalues are real and when are they positive or negative. We find that all four eigenvalues are real for $k=0$.  However, for $k \neq 0$, the condition
{\footnotesize
\begin{equation}
\beta \leq  \frac{1}{4 (\frac{\gamma}{\lambda}) \sqrt{\bar{\rho_A}\bar{\rho_B}} } \left(\frac{D}{\lambda}\frac{|k|^2}{8} + \frac{\lambda}{D}\frac{2}{|k|^2} -1 \right).
\end{equation}
}
is needed in order for $\alpha_{1,2}$ to be real, and
{\footnotesize
\begin{equation}
\beta \geq  -\frac{1}{4 (\frac{\gamma}{\lambda}) \sqrt{\bar{\rho_A}\bar{\rho_B}} } \left(\frac{D}{\lambda}\frac{|k|^2}{8} + \frac{\lambda}{D}\frac{2}{|k|^2} - 1 \right).
\end{equation}
}
is needed to ensure that $\alpha_{3,4}$ is real.  We note that the condition on $\alpha_{3,4}$ is satisfied for all $\beta \geq 0$, and $\lambda > 0$, $\gamma \geq 0$, $D\geq 0$.

The well-mixed solution is linearly stable when all four of the eigenvalues have negative real part. Starting with $\alpha_1$, the eigenvalue has a negative real part if
{\footnotesize
\begin{equation*}
\operatorname{Re} \left(4 \lambda +D |k|^2 +  \sqrt{16 \lambda^2- 8D(\lambda+4\beta \gamma \sqrt{\bar{\rho_A}\bar{\rho_B}})|k|^2 + D^2|k|^4}\right) > 0.
\end{equation*}
}
For $\lambda > 0$ and $D>0$, this is always true, and thus $\alpha_1$ always has a negative real part for the parameter values we consider.
Looking at the second eigenvalue, $\alpha_2$ has a negative real part if
{\footnotesize
\begin{equation*}
\operatorname{Re} \left( 4 \lambda +D |k|^2 -  \sqrt{16 \lambda^2- 8D(\lambda+4\beta \gamma \sqrt{\bar{\rho_A}\bar{\rho_B}})|k|^2 + D^2|k|^4} \right) > 0,
\end{equation*}
}
\noindent or equivalently,
{\footnotesize
\begin{align*}
4 \lambda +D |k|^2  &> \operatorname{Re} \left( \sqrt{16 \lambda^2- 8D(\lambda+4\beta \gamma \sqrt{\bar{\rho_A}\bar{\rho_B}})|k|^2 + D^2|k|^4} \right). \notag
\end{align*}
}
Squaring both sides and simplifying, we find
{\footnotesize
\begin{align*}
\lambda&> -2 \beta \gamma \sqrt{\bar{\rho_A}\bar{\rho_B}} \iff \beta > -\frac{1}{2(\frac{\gamma}{\lambda})\sqrt{\bar{\rho_A}\bar{\rho_B}}}.
\end{align*}
}
Hence, $\alpha_2$ also always has a negative real part for the parameter values that we consider.

Continuing on to the third eigenvalue, we recall that $\alpha_3$ is always real, so $\alpha_3$ has a negative real part if
{\footnotesize
\begin{equation*}
4 \lambda +D |k|^2 +  \sqrt{16 \lambda^2- 8D(\lambda - 4\beta \gamma \sqrt{\bar{\rho_A}\bar{\rho_B}})|k|^2 + D^2|k|^4} > 0.
\end{equation*}
}
This always holds for our choice of parameters, thus $\alpha_3$ has a negative real part for all wave numbers $k$. Finally, $\alpha_4$ is also always real for our parameter choices and has negative real part when
{\footnotesize
\begin{align*}
4 \lambda +D |k|^2 &> \sqrt{16 \lambda^2- 8D(\lambda - 4\beta \gamma \sqrt{\bar{\rho_A}\bar{\rho_B}})|k|^2 + D^2|k|^4}.
\end{align*}
}
Squaring both sides and simplifying, we find that $\alpha_4$ has negative real part exactly when
{\footnotesize
\begin{equation}
\beta < \frac{1}{2(\frac{\gamma}{\lambda})\sqrt{\bar{\rho_A}\bar{\rho_B}}}.
\end{equation}
}
Hence, $\alpha_4$ is the only eigenvalue which can have a positive real part, and the uniformly distributed solution becomes linearly unstable for
{\footnotesize
\begin{equation}
\beta \geq \frac{1}{2(\frac{\gamma}{\lambda})\sqrt{\bar{\rho_A}\bar{\rho_B}}}. \label{E:DimensionalUnstableEV}
\end{equation}
}
This allows us to define a critical $\beta$ value where the stationary solution changes stability.

\begin{figure}[!htb]
        \subfigure{
                \includegraphics[width=.44\linewidth,keepaspectratio]{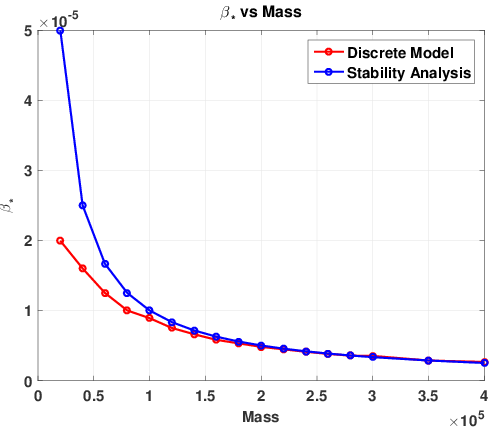}%{../Figures/f26.pdf}
						}
        \subfigure{
               \includegraphics[width=.45\linewidth,keepaspectratio]{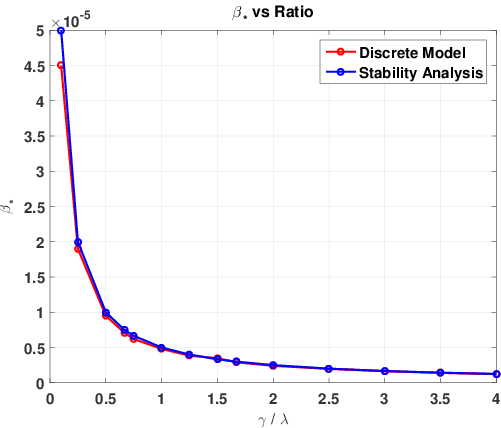}%{../Figures/f37.pdf}
               			}
        \caption[Critical $\beta$ against the system mass and ratio.]{On the Left: Critical $\beta$ against the system mass. Here we have that  $\lambda = \gamma =0.5$, and for the discrete model we have $\delta t = 1$ and the lattice size is $50 \times 50$. The red and blue curves represent the discrete model and the linearized PDE system respectively.  On the right: Critical $\beta$ against the ratio $\frac{\gamma}{\lambda}$. Here we have $N_A = N_B = 100,000$, and for the discrete model we have that $\delta t = 1$ and the lattice size is $50 \times 50$. The red and blue curves represent the discrete model and the linearized PDE system respectively.}
                \label{fig:dimensionalf52}
\end{figure}

In Fig. \ref{fig:dimensionalf52}, we plot the critical parameter value $\beta_*$ that we found numerically for the discrete model in red and the critical $\beta$ that we found in the continuum system in blue.  In the left figure, we plot them as a function of the mass $N_A + N_B$, where we fix $N_A=N_B$ and $\gamma=\lambda=.5$, and in the figure on the right, we plot them as a function of the ratio $\frac{\gamma}{\lambda}$, with $N_A=N_B=100,000$. In the figure on the left, we see that the critical $\beta_*$ from the discrete model matches the linearized PDE system's critical $\beta$ values as the system mass increases. Thus, for sufficiently large mass, the continuum equations predict our discrete model results. We also see that as the system mass increases, the phase transition occurs at a smaller $\beta$. The match between the blue and the red plots is very good except for masses below $200,000$.  However, this is not surprising, because our derivation of the continuum system is a formal derivation, where we have assumed smoothness in densities which is only achieved when there are sufficiently many agents in the system. In the figure on the right, we clearly see that the critical $\beta$ values of the discrete model very closely matches that of the linearized continuum system for all plotted ratio values. Hence, the results from our continuum equations predict our discrete model results. We also observe that as the $\frac{\gamma}{\lambda}$ ratio increases, the phase transition occurs at a smaller $\beta$.  

\section{\label{S:conclusion}Conclusions} 

In this work, we have presented an agent-based model for gang territorial development motivated by graffiti markings.  The model undergoes a phase transition as the parameters are changed.  By deriving a continuum version of the model and performing a linear stability analysis on the well-mixed state, a bifurcation point is found which matches the precise value of the critical parameter found via numerical simulations of the discrete model.

The continuum version of the model resembles a two-species version of the Keller-Segel model \cite{keller1970initiation}, though the graffiti acts as a chemo-repellent rather than a chemo-attractant, and the graffiti does not diffuse in space.  Another interesting thing to notice is that the coarsening that we observe in the simulations of the model, see for example Fig. \ref{fig:64SegF1}, closely resembles the Cahn-Hilliard equation \cite{cahn1958free}, though no connection has yet been found.  In short, there is much which remains to be explored about the models presented here.

%%------------------------------------------------
\begin{acknowledgements}
This research was supported by the National Science Foundation under Grant No. DMS-1319462. The authors are grateful to Gil Ariel, Daniel Balagu\'e Guardia, Jes\'us Rosado Linares, Wanda Strychalski, and Marie-Therese Wolfram for their helpful discussions and comments. 
\end{acknowledgements}

\begin{appendix}
\section{\label{section:dimension}Non-Dimensionalization}
In this appendix, we will derive the nondimensionalized system for the continuum equations derived in Sec. \ref{S:derivation}. We start the non-dimensionalization by first defining the natural time scale and characteristic length to be
\begin{align*}
\tilde{t} &= \lambda t, \\
\tilde{X} &= \frac{X}{l_c},
\end{align*}
\noindent where $X=(x,y)$ and $l_c = \sqrt{\frac{D}{\lambda}}$. Hence, $\tilde{t}$ and $\tilde{X}$ are dimensionless quantities. Because we are interested in the effect of variations of $\beta$ on the dynamics of the system, we also define a nondimensional $\tilde{\beta}$ by identifying $\beta_*$ as the critical $\beta$ and letting $\tilde{\beta} = \frac{\beta}{\beta_*}$. Note that $\beta_*$ carries the dimension $\frac{Space^2}{\text{\emph{Number of Individuals}}}$.

For the derivation we first start non-dimensionalizing the continuum equations for the graffiti. Recall that the dimensional version of the evolution equation for red graffiti density is
\begin{align}
\frac{\partial \xi_A}{\partial t} &= \gamma \rho_A - \lambda \xi_A. \notag \\
\intertext{Dividing both sides by  $\lambda$, we find} \notag
\frac{\partial \xi_A}{\lambda \partial t} &=  \frac{\gamma}{\lambda}\rho_A - \xi_A.  \notag \\
\intertext{Note that $\frac{\partial}{\lambda \partial t} = \frac{\partial}{\partial \tilde{t}}$, and multiplying both sides by $ \beta_* $, we arrive at} \notag
 \beta_* \frac{\partial \xi_A}{\partial \tilde t} &=  \beta_* \frac{\gamma}{\lambda}\rho_A - \beta_* \xi_A.  \notag \\
\intertext{Noting that the dimension of $\beta_*$ is $\frac{\text{space}^2}{\text{Number of Individuals}}$. Hence, $\beta_* \xi_A = \tilde{\xi_A}$ and $\beta_* \rho_A = \tilde{\rho_A}$ are dimensionless quantities, giving us the final dimensionless form of the evolution equation:}
\frac{\partial \tilde \xi_A}{\partial \tilde t} &=  c\tilde \rho_A -  \tilde \xi_A.
\end{align}
Note that here, $c=\frac{\lambda}{\gamma}$ is a dimensionless $\pi$ number.

Next, we nondimensionalize the continuum equations for the agent densities (\ref{E:agents_continuum_eqn_1}) and (\ref{E:agents_continuum_eqn_1}). Recall that the dimensional form of the evolution equations for red agent density is
\begin{align}
\frac{\partial \rho_A}{\partial t} &=  \frac{D}{4} \nabla_X \cdot \Bigg[ \nabla_X \rho_A  + 2  \beta \Big(\rho_A \nabla_X \xi_B \Big) \Bigg]. \notag \\
\intertext{Dividing both sides by $\lambda$ and employing $\beta = \beta_* \tilde{\beta}$, we see that} \notag
\frac{\partial \rho_A}{\lambda \partial t} &=  \frac{D}{4\lambda} \nabla_X \cdot \Bigg[ \nabla_X \rho_A  + 2  \tilde \beta \beta_* \Big(\rho_A \nabla_X \xi_B \Big) \Bigg]. \notag \\
\intertext{Note again that $\frac{\partial}{\lambda \partial t} = \frac{\partial}{\partial \tilde{t}}$.}
\intertext{Next, we rewrite the operator $\nabla_X$ in terms of $\tilde X$, i.e. $\nabla_X =  \sqrt{\frac{\lambda}{D}} \nabla_{\tilde X} $: } \notag
\frac{\partial \rho_A}{\partial \tilde t} &=  \frac{D}{4\lambda} \frac{\lambda}{D} \nabla_{\tilde X} \cdot \Bigg[ \nabla_{\tilde X} \rho_A  + 2  \tilde \beta \beta_*\Big(\rho_A \nabla_{\tilde X} \xi_B \Big) \Bigg]. \notag \\
\intertext{Finally, we multiply both sides by $\beta_*$:} \notag
\beta_* \frac{\partial \rho_A}{\partial \tilde t} &=  \frac{\beta_*}{4} \nabla_{\tilde X} \cdot \Bigg[ \nabla_{\tilde X}\rho_A  + 2 \tilde \beta \Big( \rho_A \nabla_{\tilde X}( \beta_* \xi_B) \Big) \Bigg], \notag
\intertext{giving us} \notag
\frac{\partial \tilde \rho_A}{\partial \tilde t} &=  \frac{1}{4} \nabla_{\tilde X} \cdot \Bigg[ \nabla_{\tilde X} \tilde \rho_A  + 2  \tilde \beta \Big(\tilde \rho_A \nabla_{\tilde X} \tilde \xi_B \Big) \Bigg].
\end{align}

Dropping the tilde notation from the equations for notational simplicity gives us the dimensionless form of the continuum system:
\begin{equation}\label{E:NonDimensionalContinuumEquations}
\begin{cases}
\frac{\partial \xi_A}{\partial t} =  c  \rho_A -  \xi_A  \\
\frac{\partial \xi_B}{\partial t} =  c  \rho_B -  \xi_B  \\
\frac{\partial \rho_A}{\partial t} =  \frac{1}{4} \nabla_{ X} \cdot \Big[  \nabla_{ X}  \rho_A  + 2  \beta  \big( \rho_A \nabla_{ X}  \xi_B \big) \Big] \\
\frac{\partial  \rho_B}{\partial \tilde t} =  \frac{1}{4} \nabla_{ X} \cdot \Big[ \nabla_{ X}  \rho_B  + 2  \beta \big( \rho_B \nabla_{ X}  \xi_A \big) \Big].
\end{cases}
\end{equation}

We see that the non-dimensional form of the continuum equations is similar to that of the dimensional form in equation (\ref{T:continuum_eqns}). However, it is sometimes more convenient to analyze the non-dimensional continuum equations, as it allows us to see how the model behaves as different parameters are scaled in relation to one another.
\end{appendix}

\bibliographystyle{plain}

\end{document}